\documentclass[10pt]{article}
\usepackage{nccmath}
\usepackage{subcaption}
\usepackage{amsmath}
\usepackage{amssymb}
\usepackage{mathrsfs}
\usepackage{graphicx}
\usepackage{cite}
\usepackage[left=2.5cm,right=2.5cm,top=1.2cm,bottom=1.3cm,includeheadfoot]{geometry}
\usepackage{indentfirst}
\usepackage{color}
\usepackage{authblk}
\usepackage{hyperref}
\usepackage[notlot,notlof,nottoc]{tocbibind}

\newcommand{\vp}{v_{+}}
\newcommand{\vm}{v_{-}}
\newcommand{\corr}{r}

\begin{document}

\begin{center}
{\Large\bf Semiclassical study of the Mixmaster model:
	The quantum Kasner map}

\vskip 4mm

David Brizuela\footnote{ david.brizuela@ehu.eus}
and
Sara F. Uria\footnote{sara.fernandezu@ehu.eus}

\vskip 3mm
{\sl Department of Physics and EHU Quantum Center, University of the Basque Country UPV/EHU,\\
	Barrio Sarriena s/n, 48940, Leioa, Spain}

\end{center}

 \date{\today}
 
 \begin{abstract}
 
According to the Belinski-Khalatnikov-Lifshitz conjecture, close to a spacelike singularity different spatial
points decouple, and the dynamics can be described in terms of the Mixmaster (vacuum Bianchi IX) model.
In order to understand the role played by quantum-gravity effects
in this context, in the present work we consider the semiclassical behavior of this model.
Classically, this system undergoes a series of transitions between Kasner epochs, which are
described by a specific transition law. This law is derived based on the conservation of certain physical quantities
and the rotational symmetry of the system. In a quantum scenario, however, fluctuations and higher-order
moments modify these quantities, and consequently also the transition rule.
In particular, we perform a canonical quantization of the model and then analytically obtain
the modifications of this transition law for semiclassical states peaked around classical trajectories.
The transition rules for the quantum moments are also obtained, and a number
of interesting properties are derived concerning the coupling between the different degrees of freedom.
More importantly, we show that, due to the presence of quantum-gravity effects and contrary to the
classical model, there appear certain finite ranges of the Kasner parameters for which the system will not undergo further transitions and will follow a Kasner regime until the singularity.
Even if a more detailed analysis is still needed,
this feature points toward a possible resolution of the classical chaotic behavior of the model.
Finally, a numerical integration of the equations of motion is also performed
in order to verify the obtained analytical results.
\end{abstract}

\section{Introduction}

According to the Belinski-Khalatnikov-Lifshitz conjecture\cite{BKL}, close to a generic spacelike singularity,
spatial derivatives become negligible as compared to time derivatives. The dynamics of the spacetime becomes then effectively local, in the sense that different spatial points decouple,
and in vacuum the evolution of each point can thus be described in terms of the Mixmaster
(vacuum Bianchi IX) model \cite{Misner,Misner2}.
Even if complicated, the dynamics of this model is quite well understood, and asymptotically,
as the singularity is approached, it can be described in terms of a sequence of Bianchi II models
combined with appropriate rotations of the axes \cite{Heinzle:2009eh,Ringstrom}.
This Bianchi II model in turn predicts kinetically dominated long periods, which are known
as Kasner epochs, connected by quick transitions.
In general, while the spatial volume monotonically decreases as the singularity is approached,
the scale factors corresponding to the different directions oscillate, periodically stretching and squeezing, which results in chaotic behavior of the model.

Although a formal proof is still missing, this conjecture is supported by a number of numerical studies \cite{Berger:2002st,Garfinkle:2003bb,Heinzle:2012um}. However, it cannot be complete, since near any singularity, quantum-gravity effects are expected to become important. Hence, in order to study how this picture is modified when the quantum behavior of the gravity is taken into account, one needs to consider the quantum dynamics of the Mixmaster model.

In the literature numerous approaches can be found that study such models based on various quantization methods and focused on different questions, such as the avoidance of the singularity, the survival of the chaotic behavior, or the construction of fundamental and excited quantum states~\cite{Berger:1989jm,Bojowald:2004ra,Kheyfets:2005nz,Benini:2006xu,Wilson2010,Ashtekar:2011ck,Bergeron:2015ppa,Czuchry:2016rlo,Wilson-Ewing:2018lyx,Kiefer:2018uyv,Gozdz:2018aai,Giovannetti:2019ewe,Bae,Bergeron:2015lka,Bergeron:2017gte}. Other proposals focus on Bianchi II quantum models~\cite{Ashtekar:2009um,Corichi:2012hy,Bergeron:2014kea,Saini:2017ipg,Ours:LRS} as a simpler setup to provide a preliminary description of the full quantum evolution.
In particular, an interesting question is whether the structure of alternating Kasner regimes and transitions endures when quantum effects are considered and, more importantly, how the classical transition laws
are modified. This was first examined in Ref.~\cite{Ours:LRS} for the locally rotationally symmetric Bianchi II model. In the present paper we will generalize this study for a general vacuum Bianchi IX universe. 
For such a purpose, we will first analyze the canonical structure and the corresponding approximate solutions of the classical model. In this way, we will obtain the transition laws that relate the characteristic parameters
of subsequent Kasner regimes, also known as the Kasner map. In order to perform the analysis of the quantum system, the wave function will be decomposed into its
infinite set of moments by following the framework first presented in Ref. \cite{Bojowald:2005cw}. This framework is especially well suited to study the evolution of peaked semiclassical states,
and it has proven to be very useful to describe the quantum dynamics of several cosmological models see, e.g., Refs. \cite{Bojowald:2020emy,Baytas:2018gbu,Baytas:2016cbs,Bojowald:2015fla,Brizuela:2015cna,Brizuela:2014dda,Brizuela:2014cfa,Bojowald:2014qha,Bojowald:2010qm}. In this context, with certain approximations,
we will analytically solve the dynamics of the system in order
to check how the classical transition laws are modified by quantum-gravity effects. Furthermore, we will
also obtain the relation between quantum moments of subsequent Kasner regimes. This study will lead to the complete quantum Kasner map.
In particular we will show that, contrary to the classical map, in this case
there are finite ranges of the Kasner parameters for which the system does not develop further transitions,
and thus follows a Kasner regime until the singularity. This fact indicates that, at least in this sense,
the stochastic properties of the classical model will not be present in the quantum case.
Finally, we will also perform a numerical resolution of the full
equations of motion in order to test our analytical results.

The rest of the article is organized as follows: In Sec. \ref{sec:classical}, we perform the classical analysis of
the model. Section \ref{sec:quant} presents the analytical study of the quantum dynamics and, in particular,
the quantum Kasner map for a generic state. In Sec. \ref{sec:specific_state}, certain particular initial quantum states
are considered in order to obtain some specific features of the corresponding quantum Kasner map
and to perform a numerical study of the system. Section \ref{sec:conclusions} discusses and summarizes the main results of the paper.

\section{Classical analysis}
\label{sec:classical}
\subsection{Variables and equations of motion}
\label{subsec:classi_variables_eqm}
There is a wide variety of variables used in the literature to describe the Bianchi IX model.
Here, we will follow the framework proposed by Misner~\cite{Misner2}, who defined the shape parameters:
\begin{align}\label{b_plus_b_minus}
\begin{split}
\beta_+&:=-\frac{1}{2}\ln
\Bigg[\frac{a_3}{(a_1a_2a_3)^{1/3}}\Bigg], 
 \\[7pt]
\beta_-&:=
\frac{1}{2\sqrt{3}}\ln
\bigg(\frac{a_1}{a_2}\bigg),
\end{split}
\end{align}
with $a_1$, $a_2$, and $a_3$ being the scale factors for each spatial direction.
These shape parameters are vanishing in the case of a completely isotropic universe $(a_1=a_2=a_3)$
and provide a direct measure of the anisotropicity of the model:
$\beta_-$ encodes the ratio between the scale factors in directions 1 and
2, whereas $\beta_+$ represents the relation between the scale factor in direction
3 and (the cube root of) the volume $e^\alpha:=(a_1 a_2 a_3)^{1/3}$.

In the previous paper \cite{Ours:LRS}, we considered the particular locally rotationally symmetric case $a_1=a_2$,
which removes the shape parameter $\beta_-$ and greatly simplifies the computations.
But here, we will analyze the general case described by the three configuration variables $(\alpha, \beta_+, \beta_-)$.

The Hamiltonian constraint for the vacuum  Bianchi IX model reads as follows:
\begin{equation}\label{constraint}
{\cal C}=\frac{1}{2} e^{-3 \alpha}
\left(-p_\alpha^2+p_{-}^2+p_{+}^2\right)+e^{\alpha} U(\beta_+,\beta_-)=0,
\end{equation}
where $p_{\alpha}:=-\frac{e^{3\alpha}}{N} \frac{d\alpha}{dt}$ and $p_{\pm}:=\frac{e^{3\alpha}}{N} \frac{d\beta_{\pm}}{dt}$
are the conjugate momenta of $\alpha$ and $\beta_{\pm}$,  respectively, with the lapse function $N$ and corresponding
time variable $t$. The Poisson brackets are canonical and are given by 
$\{\alpha, p_\alpha \}=\{\beta_+, p_+\}=\{\beta_-, p_-\}=1$, and equal zero for any other combination of variables.
Since $\alpha=\alpha(t)$ is a monotonic function, and as it is usually done in this model,
we will choose $\alpha$ as the internal time variable.
However, it is important to keep in mind that different choices of deparametrization are possible,
which generically would give rise to different quantum pictures. In this way, one can deparametrize the above
constraint and construct the physical (nonvanishing) Hamiltonian,
\begin{align}\label{hamiltonian}
	H:=&-p_\alpha=\big[p_+^2+p_-^2+2e^{4\alpha}U(\beta_+,\beta_-)\big]^{1/2}.
\end{align}
Then, the evolution equations are straightforwardly obtained by
computing the Poisson brackets of the corresponding
variable with the Hamiltonian 
 [Eq.~\eqref{hamiltonian}]---namely,
\begin{eqnarray}
\label{eq_m_beta_+}
\dot{\beta}_{+}&=& \frac{p_+}{H},
\\
\label{eq_m_beta_-}
\dot{\beta}_{-}&=&
\frac{p_-}{H},
\\
\label{eq_m_p_+}
\dot{p}_{+}&=&-\frac{e^{4 \alpha }}{H}\frac{\partial U}{\partial\beta_+},
\\
\label{eq_m_p_-}
\dot{p}_{-}&=&-\frac{e^{4 \alpha }}{H}\frac{\partial U}{\partial\beta_-},
\end{eqnarray}
where the dot stands for a derivative with respect to $\alpha$. 

\begin{figure}[h!]
	\centering
	\includegraphics[width=0.5\linewidth]{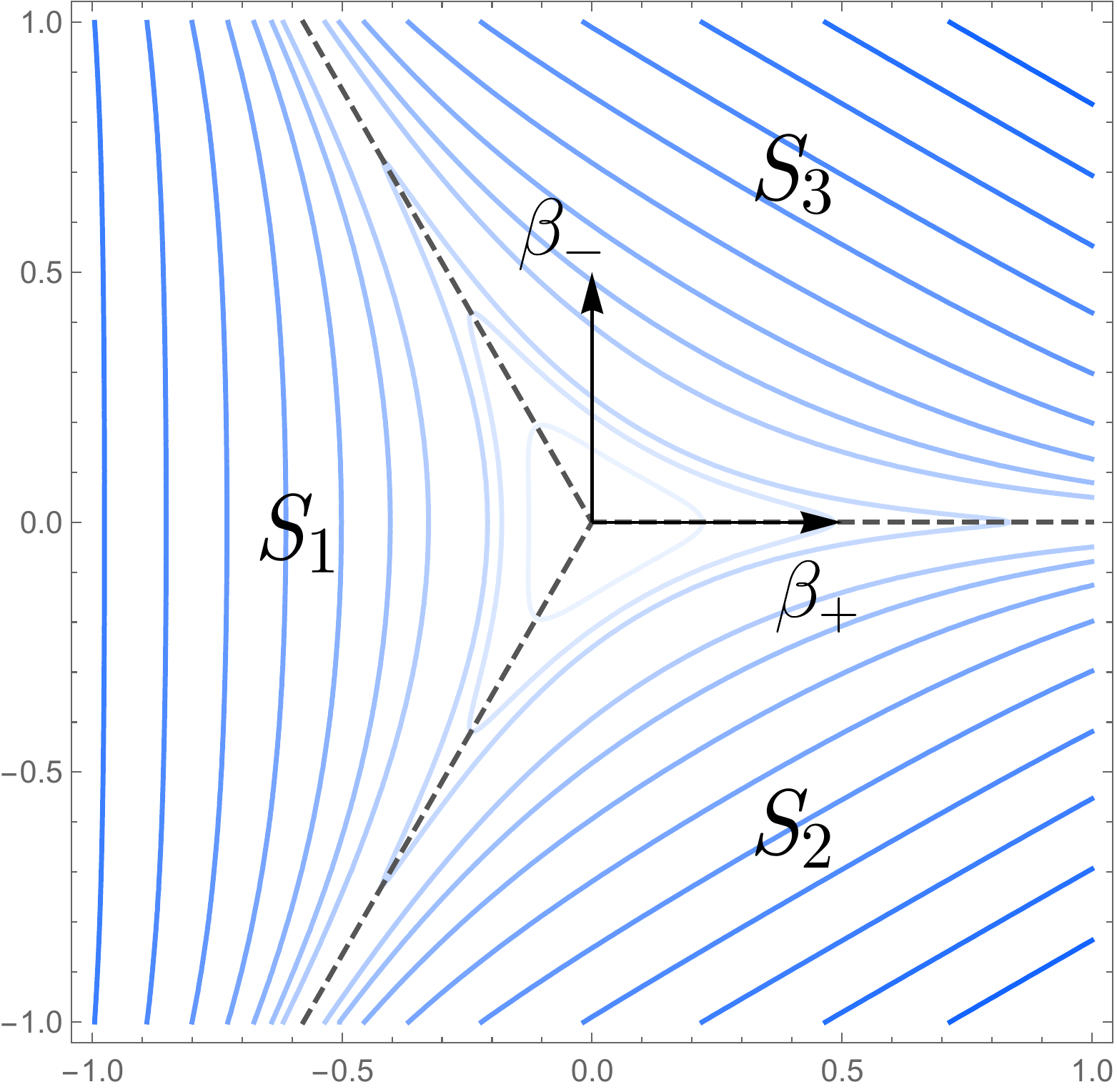}
	\caption{Equipotential plot of the Bianchi IX potential $U(\beta_+,\beta_-)$ given in Eq. \eqref{potential_Bianchi_IX}.
	The continuous blue curves are the equipotential lines, with the brightness denoting the value of the
	potential (brighter lines correspond to higher values of the potential).
	The three dashed black lines describe the local minima of the potential. Along these lines, the potential takes
	negative values---from the global minimum at the origin, $U(\beta_+,\beta_-)=-1/2$, to $U(\beta_+,\beta_-)\rightarrow 0$ as the
	shape parameters tend to infinity.}
	\label{fig:equipotential}
\end{figure}

The potential $U$ only depends on
the shape parameters and has the explicit form
\begin{align}\label{potential_Bianchi_IX}
&U(\beta_+,\beta_-):=\frac{1}{6}\left[
e^{-8\beta_+}+2e^{4\beta_+}\left(
\cosh(4\sqrt{3}\beta_-)-1
\right)-4e^{-2\beta_+}\cosh(2\sqrt{3}\beta_-)
\right].
\end{align}
The equipotential plot of this function is shown in Fig.~\ref{fig:equipotential}.
This potential has a threefold rotational symmetry with respect to the origin---that is, for any integer $n$, it is invariant under a $2\pi n/3$ rotation,
\begin{equation}
 U(\beta_+,\beta_-)= U(O^n(\beta_+,\beta_-)),
\end{equation}
where $O$ describes a $2\pi/3$ clockwise rotation. Its explicit action is given by
$O(x,y):=\left(-\frac{x}{2}+\frac{\sqrt{3}y}{2},-\frac{\sqrt{3}x}{2}-\frac{y}{2}\right)$,
which can easily be seen from its matrix representation,
\begin{align}\label{rotation}
\begin{pmatrix}
\cos\left(\frac{2\pi}{3}\right) & \sin\left(\frac{2\pi}{3}\right) \\
-\sin\left(\frac{2\pi}{3}\right) & \cos\left(\frac{2\pi}{3}\right) 
\end{pmatrix}=
\begin{pmatrix}
-\frac{1}{2} & \frac{\sqrt{3}}{2} \\
-\frac{\sqrt{3}}{2} & -\frac{1}{2} 
\end{pmatrix}.
\end{align}
In fact, since the kinetic term is invariant under rotations,
the whole Hamiltonian [Eq.~\eqref{hamiltonian}] has this rotational symmetry
considering the transformation of the four canonical variables $(\beta_+,\beta_-,p_+,p_-)$---that is,
\begin{align}\label{rot_hamiltonian}
	 H(\beta_+,\beta_-,p_+,p_-,\alpha)= H(R^n(\beta_+,\beta_-,p_+,p_-),\alpha),
\end{align}
where $R$
denotes the canonical transformation produced
by a $2\pi/3$ clockwise rotation both in the position variables and its corresponding momenta,
\begin{align}\label{C_4_dim}
R(\beta_+,\beta_-,p_+,p_-):=\left(-\frac{1}{2}\beta_++\frac{\sqrt{3}}{2}\beta_-,-
\frac{\sqrt{3}}{2}\beta_+-\frac{1}{2}\beta_-,-\frac{1}{2}p_++\frac{\sqrt{3}}{2}p_-,
-
\frac{\sqrt{3}}{2}p_+-\frac{1}{2}p_-\right).
\end{align}
An important property of this transformation, that will later be used, is that $R^{2}=R^{-1}$,
since two $2\pi/3$ clockwise rotations are equivalent to a $2\pi/3$ counterclockwise rotation.
In fact, the full symmetry group of the Hamiltonian is $C_{3v}$, which is generated by the $2\pi/3$ rotation and a reflection relative to one the symmetry semiaxes.

As can be seen in Fig. \ref{fig:equipotential},
the three symmetry semiaxes $\{\beta_-=0, \beta_+>0\}$, $\{\beta_-=\sqrt{3}\beta_+/2, \beta_->0\}$, and
$\{\beta_-=-\sqrt{3}\beta_+/2, \beta_-<0\}$ define three valleys surrounded by exponential walls
and divide the $(\beta_+,\beta_-)$ plane into three sectors,
which will be named $S_1$, $S_2$, and $S_3$.
All along these semiaxes the potential takes negative values and runs from the global minimum $U(0,0)=-1/2$ at the origin
toward zero as the shape parameters tend to infinity. Away from these valleys, the potential can be well approximated in terms of the much
simpler exponential function $V(\beta_+,\beta_-)=e^{-8\beta_+}/6$, which corresponds to the Bianchi II model
and appropriately chosen rotations. More precisely,
\begin{align}U(\beta_+,\beta_-)\approx\begin{cases}
\label{pot_1}
V(\beta_+,\beta_-),
\quad &\text{if $(\beta_+,\beta_-)\in S_1$},
\\
(V\circ O)(\beta_+,\beta_-)=V\left(-\frac{\beta_+}{2}-\frac{\sqrt{3}\beta_-}{2},\frac{\sqrt{3}\beta_+}{2}-\frac{\beta_-}{2}\right),\quad &\text{if $(\beta_+,\beta_-)\in S_2$},
\\
(V\circ O^{-1})(\beta_+,\beta_-)=V\left(-\frac{\beta_+}{2}+\frac{\sqrt{3}\beta_-}{2},-\frac{\sqrt{3}\beta_+}{2}-\frac{\beta_-}{2}\right),\quad &\text{if $(\beta_+,\beta_-)\in S_3$},
\end{cases}
\end{align}
where the symbol $\circ$ stands for the composition of the corresponding functions.
That is, away from the valleys, in sector $S_1$, the potential is directly given by $V$.
In sectors $S_2$ and $S_3$, the leading terms of the potential $U$ are, respectively,
$e^{4(\beta_+-\sqrt{3}\beta_-)}$ and $e^{4(\beta_++\sqrt{3}\beta_-)}$, and thus one needs to perform a $2\pi/3$
rotation to approximate them
by the simpler function $V$.

In particular, all the above properties of the potential $U$ make it possible
to asymptotically (as the Universe approaches the curvature singularity located
at $\alpha\rightarrow-\infty$) describe the dynamics of a generic Bianchi IX model
as a sequence of Kasner epochs and transitions ruled by the Bianchi II potential $V$
\cite{Ringstrom}. During the Kasner epochs
the potential $U$ is negligible, the kinetic term of the Hamiltonian thus dominates,
and the system follows a straight line in the $(\beta_+,\beta_-)$ plane.
At a certain point, the system ceases to be kinetically dominated, it bounces against
the exponential walls described by the potential $U$, which can be approximated
by the Bianchi II potential $V$, and it enters a new Kasner epoch.

Therefore, in the following subsections we will derive 
the transition rules (sometimes also called the Kasner map)
that relate the precise values of the characteristic constants of subsequent Kasner epochs.
More precisely, in Sec. \ref{subsec:kasner} we will first solve
the equations of motion for a given Kasner epoch and characterize each of these
epochs by four constants. Then, the analysis of the transition rules will be divided
into two steps. In Sec. \ref{subsec:bounce} a bounce of the system against the exponential
walls will be considered, whereas in Sec. \ref{subsec:sequence_bounces} this study
will be extended to describe a sequence of bounces against the exponential walls.

\subsection{Kasner epochs}\label{subsec:kasner}

The dynamics of the exact Kasner (Bianchi I vacuum) model is ruled by the Hamiltonian in Eq. \eqref{hamiltonian}
with a vanishing potential---that is, $ H_K=P:=(p_+^2+p_-^2)^{1/2}$, which corresponds
to the square root of the Hamiltonian of a free particle in two dimensions.
Thus, it does not depend on the shape parameters, and in this case
the momenta $p_{+}$ and $p_-$ remain constant all along the evolution.
As commented above, asymptotically the Bianchi IX model follows certain regimes where the dynamics can be well approximated by the Hamiltonian of the Kasner model $H_K$. However,
before explicitly solving the equations of motion for these regimes, let us first analyze the conditions
that must be met for the potential and its derivatives for such a regime to be a good approximation of
the full Bianchi IX dynamics.

First of all, the potential $U$ must be negligible with respect to the kinetic term $P$---that is,
\begin{align}
	\label{kasner_cond_1}
	\frac{e^{4\alpha}|U|}{P^2}\ll 1,
\end{align}
 where $P=(p_+^2+p_-^2)^{1/2}$ as defined above. In addition, the derivatives of the potential
should also be small, since otherwise, due to the equations of motion [Eqs.~\eqref{eq_m_beta_+}--\eqref{eq_m_p_-}],
the momenta will not be constant and the Kasner regime will not be followed. In order to check
the exact requirements for the gradient of the potential, one can perform a Taylor expansion
of the functions $p_{\pm}(\alpha)$ around a certain value of $\alpha$. Then, for these momenta to be constant
during a lapse of time $\delta\alpha$, the following terms should be negligible:
\begin{align*}
\begin{split}
\frac{|\delta\alpha|^n}{n!}
\left|\frac{1}{p_{+}}
\frac{d^np_{+}}{d\alpha^n}
\right|\ll 1,
\\
\frac{|\delta\alpha|^n}{n!}
\left|\frac{1}{p_{-}}
\frac{d^np_{-}}{d\alpha^n}
\right|\ll 1,
\end{split}
\end{align*}
for all integers $n\geq 1$.
By combining these expressions with the equations of motion [Eqs. \eqref{eq_m_p_+} and \eqref{eq_m_p_-}], we obtain the precise conditions for the derivatives of the potential:
\begin{align}\label{kasner_cond_2}
&e^{4\alpha}
\frac{|\delta\alpha|^n}{n!}
\left|
\gamma_n^{+}\right|
\ll 1,
\\
\label{kasner_cond_3}
&e^{4\alpha}
\frac{|\delta\alpha|^n}{n!}
\left|
\gamma_n^{-}\right|
\ll 1,
\end{align}
for all integers $n\geq 1$, and where $\gamma_n^{\pm}$ have been defined as
\begin{align}\label{kasner_cond_a_n}
	&\gamma_n^+:=\frac{1}{P^n}\sum_{k=1}^n\binom{n-1}{k-1}p_+^{k-2}p_-^{n-k}
	\frac{\partial^n U}{\partial \beta_{+}^k\partial\beta_-^{n-k}},
	\\
	&\gamma_n^-:=\frac{1}{P^n}\sum_{k=1}^n\binom{n-1}{k-1}p_-^{k-2}p_+^{n-k}
	\frac{\partial^n U}{\partial \beta_{-}^k\partial\beta_+^{n-k}}.
\end{align} 
The equations of motion for the shape parameters in Eqs. \eqref{eq_m_beta_+} and \eqref{eq_m_beta_-} do not provide extra requirements. Therefore, as long as the conditions of Eqs.
\eqref{kasner_cond_1}--\eqref{kasner_cond_3} are obeyed at a given value of $\alpha$,
the Kasner regime will be a good approximation of the dynamics during a lapse of time $\delta\alpha$.
In fact, from these conditions one can estimate the remaining time $\delta\alpha$ until the next bounce as
\begin{align}\label{key}
	|\delta\alpha|\approx e^{-4\alpha}
	\min_{n\geq 1}\left\{\left(
	\frac{n!}{\left|
		\gamma_n^{+}\right|}\right)^{1/n},
	\left(
	\frac{n!}{\left|
		\gamma_n^{-}\right|}\right)^{1/n}
	\right\}.
\end{align}
As can be observed, due to the exponential term $e^{-4\alpha}$, the duration of the Kasner epochs grows as $\alpha\to -\infty$. Consequently, as the system approaches the singularity, each subsequent
Kasner regime lasts longer than the previous ones. 

As already commented,
during these Kasner periods, the momenta $p_+$ and $p_-$ are conserved,
and the corresponding equations of motion for the shape parameters
[Eqs. \eqref{eq_m_beta_+} and \eqref{eq_m_beta_-}] can easily be solved:
\begin{eqnarray}\label{beta_+_kas}
\beta_+&=&c_++\dfrac{p_+}
{P}\,\alpha,
\\
\label{beta_-_kas}
\beta_-&=&c_-+\dfrac{p_-}
{P}\,\alpha,
\end{eqnarray}
with the integration constants $c_{+}$ and $c_-$. If we consider the system as
a particle moving along the plane $(\beta_+,\beta_-)$,
then since we are considering the backward evolution as we
approach the singularity at $\alpha\rightarrow-\infty$,
it is natural to define the velocity vector as $\mathbf{v}:=(-\dot{\beta}_+,-\dot{\beta}_-)$.
According to Eqs. \eqref{beta_+_kas} and \eqref{beta_-_kas}, during the Kasner regimes this vector $\mathbf{v}$
is constant and of unit Euclidean norm. Therefore, in this sense, all
the dynamical information can be encoded in the angle $\theta$
formed by $\mathbf{v}$ with the positive $\beta_+$ axis. That is, defining the angle $\theta\in[0,2\pi)$ as $\cos\theta=-\frac{p_+}{P}
$ and $\sin\theta=-\frac{p_-}{P}
$, the above expressions take the simple form
\begin{eqnarray}\label{beta_+_kas_theta}
\beta_+&=&c_+ -\alpha \cos\theta,
\\
\label{beta_-_kas_theta}
\beta_-&=&c_- - \alpha\sin\theta.
\end{eqnarray}
Hence, during these Kasner epochs the shape parameters are, depending on the sign
of their corresponding momentum $p_-$ or $p_+$, increasing or decreasing linear functions
of $\alpha$,  and the system follows a straight line in the $(\beta_+,\beta_-)$ plane
with slope $\tan\theta=\frac{p_-}{p_+}$. In this way, each Kasner epoch is completely characterized
by the values of the four parameters $(c_+,c_-,p_+,p_-)$. These parameters are constant
during each epoch, but they will change when the system bounces against the potential wall.

\subsection{A bounce against the potential walls}\label{subsec:bounce}

In order to analyze the bounce of the system against the potential walls, it is necessary to
solve the dynamics given by the full evolution equations [Eqs. \eqref{eq_m_beta_+}--\eqref{eq_m_p_-}]. We will obtain an implicit
solution by constructing the conserved quantities $C=C(\alpha,\beta_+,p_+,\beta_-,p_-)$.
By definition, such quantities must obey
\begin{equation}
\label{const_motion_1}
0=\frac{d C}{d\alpha}=
\frac{\partial C}{\partial\alpha}
+\frac{\partial C}{\partial\beta_+}\frac{d\beta_+}{d\alpha}
+\frac{\partial C}{\partial\beta_-}\frac{d\beta_-}{d\alpha}
+\frac{\partial C}{\partial p_+}\frac{dp_+}{d\alpha}
+\frac{\partial C}{\partial p_-}\frac{d p_-}{d\alpha},
\end{equation}
which, using the equations of motion \eqref{eq_m_beta_+}-\eqref{eq_m_p_-}, can be rewritten as,
\begin{equation}
0=H\frac{\partial C}{\partial\alpha}
+\frac{\partial C}{\partial\beta_+} p_++
\frac{\partial C}{\partial\beta_-} p_-
-e^{4\alpha}\left(\frac{\partial U}{\partial\beta_+}\frac{\partial C}{\partial p_+}
+\frac{\partial U}{\partial\beta_-}\frac{\partial C}{\partial p_-} \right).
\label{const_motion_2}
\end{equation}
Due to the complicated form of the potential $U$, it turns out to be very difficult
to obtain analytical solutions to this equation. However, as explained in the beginning
of this section [see Eq. \eqref{pot_1}], for sufficiently large (in absolute value) negative $\alpha$,
the potential $U$ can be well approximated by the simpler potential $V=e^{-8\beta_+}/6$ and certain rotations.
For simplicity,
let us first consider that the bounce happens in the sector $S_1$, where $U\approx V$. (The analysis for the bounces in the other two sectors, $S_2$ and $S_3$,
will require certain rotations and will be considered below.)

If the potential $U$ is given by $V$, the above equations for the conserved quantities are simplified
since $\partial V/\partial\beta_-=0$ and $\partial V/\partial\beta_+=-8 V$. Furthermore, considering
the change of variables $\alpha\rightarrow H$, so that now $C=C(H,\beta_+,\beta_-,p_+,p_-)$,
Eq. \eqref{const_motion_2} takes the form
\begin{equation}
0=2 (H^2-p_+^2-p_-^2)\frac{\partial C}{\partial H}
+\frac{\partial C}{\partial\beta_+} p_++
\frac{\partial C}{\partial\beta_-} p_-
+4(H^2-p_+^2-p_-^2)\frac{\partial C}{\partial p_+},
\end{equation}
which is a linear differential equation with polynomial coefficients. This can be
analytically solved, which leads to the general form for the conserved quantities,
\begin{align}
C=&C\left[p_-,2H-p_+,
\left(\frac{2H-p_+}{p_-}\right)\beta_-+\beta_+-2\alpha,
e^{\frac{4[(2H-p_+)^2-3p_-^2]^{1/2}\beta_-}{p_-}}\left(
\frac{2}{1+\frac{2p_+-H}{[(2H-p_+)^2-3p_-^2]^{1/2}}}-1
\right)
\right],
\nonumber
\\\label{conserved_quant}
\end{align}
for any arbitrary function $C:\mathbb{R}^4\to\mathbb{R}$. Therefore, by choosing appropriate forms for this function,
we define the following four observables:
\begin{eqnarray}\label{conserved_quantities}
\begin{split}
C_1&:=p_-,
\\
C_2&:=2H-p_+,
\\
C_3&:=\left(\frac{2H-p_+}{p_-}\right)\beta_-+\beta_+-2\alpha,
\\
C_4&:=e^{\frac{4[(2H-p_+)^2-3p_-^2]^{1/2}\beta_-}{p_-}}\left(
\frac{2}{1+\frac{2p_+-H}{[(2H-p_+)^2-3p_-^2]^{1/2}}}-1
\right).
\end{split}
\end{eqnarray}

These observables provide the analytic solution to the system implicitly. As they are conserved
during the evolution, both during the Kasner epoch as well as during the bounce against the potential
walls, one can obtain the transition rules that relate two subsequent Kasner epochs simply by
evaluating these objects at each Kasner epoch and requesting that they indeed have the same value.

If we denote the parameters of the prebounce Kasner epoch by a bar $(\bar c_+,\bar c_-, \bar p_+,\bar p_-)$
and those of the postbounce Kasner epoch by a tilde $(\tilde c_+,\tilde c_-, \tilde p_+,\tilde p_-)$, the
relation between both is then given by
\begin{align}
\label{transition_p_-}
&\widetilde{p}_-=\overline p_-,
\\
\label{transition_p_+}
&\widetilde{p}_+=\frac{1}{3}\left(4\overline P-5\overline p_+\right)=:B_p(\overline p_+,\overline p_-),
\\
&\widetilde{c}_+=-\frac{3\overline P}{5\overline P-4\overline p_+} \overline{c}_+-\frac{2\overline P-\overline p_+}{2\left(5\overline P-4\overline p_+\right)}\ln\left[\frac{2}{3}\left(2\overline p_+-\overline P\right)^2\right]=:B_+(\overline c_+,\overline p_+,\overline p_-),
\label{transition_c_+}
\\
&\widetilde{c}_-=\overline c_-+\frac{4\overline p_-}{5\overline P-4\overline p_+} \overline{c}_++\frac{\overline p_-}{2\left(5\overline P-4\overline p_+\right)}\ln\left[\frac{2}{3}\Big(2\overline p_+-\overline P\Big)^2\right]=:B_-(\overline c_+,\overline c_-,\overline p_+,\overline p_-),
\label{transition_c_-}
\end{align}
where $\overline P:=(\overline p_+^2+\overline p_-^2)^{1/2}$. Therefore, the transition rule
$B$ that describes the bounce against the potential walls in sector $S_1$ is given as follows,
\begin{align}\label{bounce}
B(c_+,c_-,p_+,p_-)=\left(B_+(c_+,p_+,p_-),B_-(c_+,c_-,p_+,p_-),B_p(p_+,p_-),p_-\right),
\end{align}
where $B_+$, $B_-$, and $B_p$ have been defined in Eqs. \eqref{transition_p_+}--\eqref{transition_c_-}.

In order to analyze the bounces that happen in the other sectors,
it is important to first note that, due to the Kasner
dynamics [Eqs. \eqref{beta_+_kas_theta} and \eqref{beta_-_kas_theta}],
it is the prebounce angle $\theta$, or equivalently the ratio $p_-/p_+$, that is the key parameter
that will determine in which sector the bounce will take place:
namely, in $S_1$ for $\theta\in(2\pi/3,4\pi/3)$,
in $S_2$ for $\theta\in(4\pi/3,2\pi)$, and in $S_3$ for $\theta\in(0,2\pi/3)$.
As commented above, in sectors $S_2$ and $S_3$ the leading-order terms in the potential
are $e^{4(\beta_+-\sqrt{3}\beta_-)}$ and $e^{4(\beta_++\sqrt{3}\beta_-)}$, respectively, and thus
one needs to implement the corresponding rotation as given in Eq. \eqref{pot_1}.
Such rotations leave the Hamiltonian as the one corresponding to the Bianchi II model
and thus, in this rotated frame,
one can directly apply the above transition rule $B$. The rotation should then be inverted in order to provide the result in the original basis.
This leads to the explicit form of the Kasner map,
\begin{align}\label{trans_IX}
T(c_+,c_-,p_+,p_-)=\begin{cases}
B(c_+,c_-,p_+,p_-), &\theta\in \left(
\frac{2\pi}{3},\frac{4\pi}{3}
\right),
\\
R^{-1}\circ B\circ R(c_+,c_-,p_+,p_-), &\theta\in \left(
\frac{4\pi}{3},2\pi\right),
\\
R\circ B\circ R^{-1}(c_+,c_-,p_+,p_-), &\theta\in \left(0,
\frac{2\pi}{3}\right),
\end{cases}
\end{align}
with $R$ being the $2\pi/3$ clockwise rotation [Eq. \eqref{C_4_dim}].
The boundary values $\theta=0, 2\pi/3,$ and $4 \pi/3$ are not included in this map, since
for these cases, there is no bounce and the system will follow a Kasner regime until the singularity.
We will call these values \emph{exit channels} (or \emph{exit points} in the case
that they are just a set of isolated points) since the system exits the bouncing behavior.

\begin{figure}[h]
	\centering
	\includegraphics[width=0.9\linewidth]{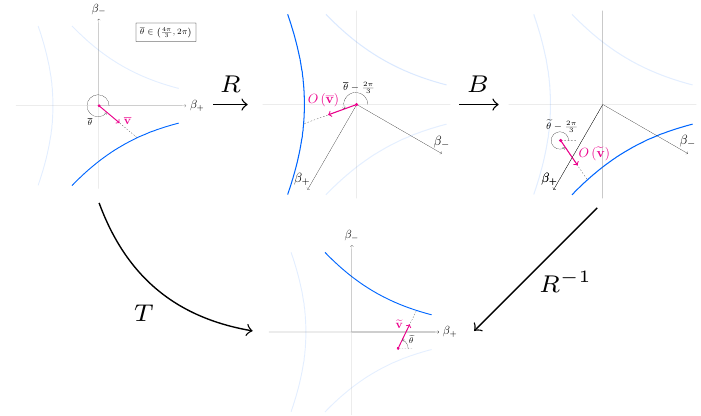}
	\caption{Schematic diagram of the computation of the Kasner map $T$ for a bounce of the system against the potential walls in sector $S_2$---that is, for a prebounce angle $\overline\theta\in\left(\frac{4\pi}{3},2\pi\right)$. In the diagram, $\mathbf{\overline v}$ and $\mathbf{\widetilde v}$
	refer to the velocity vector in the prebounce and postbounce Kasner regimes, respectively, whereas
	$O(\mathbf{\overline v})$ and $O(\mathbf{\widetilde v})$ stand for their corresponding $2\pi/3$
	clockwise rotations.}
	\label{fig:trans_IX}
\end{figure}

This transition rule is the most relevant result of this section and
can also be written in the more compact form
\begin{equation}\label{trans_IX_simp}
T(c_+,c_-,p_+,p_-)=R^{-m(\theta)}\circ B\circ R^{m(\theta)}(c_+,c_-,p_+,p_-),
\end{equation}
where the integer $m(\theta)$ is defined as $m(\theta):=\left\lfloor3\theta/2\pi-1\right\rfloor$,
and it just takes the value $m(\theta)=0$ for $\theta\in [2\pi/3,4\pi/3)$, 
$m(\theta)=1$ for $\theta\in[4\pi/3,2\pi)$,
and $m(\theta)=-1$ for $\theta\in [0,2\pi/3)$. As a particular example, a
schematic diagram of the computation of a transition that takes place in sector $S_2$
is illustrated in Fig. \ref{fig:trans_IX}. 

With this relation at hand, it is also interesting to obtain the transition rule that relates the angle of the velocity vector of the prebounce Kasner epoch $\bar\theta$ with that of the postbounce
Kasner epoch $\tilde\theta$, which reads
\begin{align}\label{transition_theta_general}
\begin{split}
\cos\widetilde\theta=-\dfrac{4\cos\left(
	\frac{2 m(\overline\theta)\pi}{3}
	\right)+4
	\cos\left(\overline\theta+
	\frac{2m(\overline\theta)\pi}{3}
	\right)
	+\cos\overline\theta}{5+4\cos\left(\overline\theta-
	\frac{2m(\overline\theta)\pi}{3}
	\right)},
\\
\sin\widetilde\theta=-\frac{4\sin\left(
	\frac{2m(\overline\theta)\pi}{3}
	\right)-4
	\sin\left(\overline\theta+
	\frac{2m(\overline\theta)\pi}{3}
	\right)
	+\sin\overline\theta}{5+4\cos\left(\overline\theta-
	\frac{2m(\overline\theta)\pi}{3}
	\right)}.
	\end{split}
\end{align}
Furthermore,
these trigonometric functions can be parametrized in terms of a single real parameter $u$---see, e.g., Ref. \cite{Belinski}
\footnote{Due to the symmetry of the model, in the literature the parameter $u$
is usually assumed to be defined for $u>1$,
which corresponds to the wedge $\theta\in(\pi,4 \pi/3)$.
The information regarding the rest of the plane is then encoded elsewhere.
However, in order to have the complete information of the dynamics of the
particle, and in particular the sector where the corresponding bounce happens,
here we will consider $u$ to be defined in the whole real line. The relation
between these two parametrizations and their corresponding transition laws
is explained in Appendix \ref{app:u_parameter}.}
:
\begin{align}\label{parametrization_u}
&\cos\theta=-
\frac{1}{2}\left(\frac{1+4u+u^2}{1+u+u^2}\right)=:c(u),
\qquad
\sin\theta=
\frac{\sqrt{3}}{2}
\left(\frac{1-u^2}{1+u+u^2}\right):=s(u),
\end{align}
with the inverse,
	\begin{equation}\label{utheta}
	u=-\frac{1+2\cos\theta}{2+\cos\theta+\sqrt{3}\sin\theta}.
	\end{equation}
\begin{figure}[h!]
	\centering
	\includegraphics[width=0.45\linewidth]{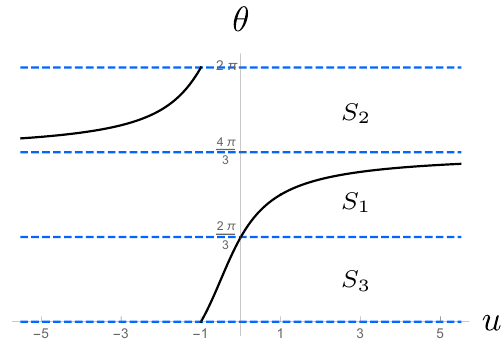}
	\caption{The black continuous curve shows the relation
		between the angle $\theta$ and the parameter $u$. The dashed horizontal
		blue lines are the boundaries between the different sectors $S_1$, $S_2$, and $S_3$.}
	\label{fig:theta_u}
\end{figure}
Therefore, as explicitly shown in Fig. \ref{fig:theta_u},
the different sectors can also be defined in terms of the value of $u$---namely,
\begin{align}\label{sectors_u}
\theta\in\begin{cases}
\left(
\frac{2\pi}{3},\frac{4\pi}{3}	\right), &u>0,
\\
\left(
\frac{4\pi}{3},2\pi\right), &u<-1,
\\
\left(0,
\frac{2\pi}{3}\right), &u\in(-1,0).
\end{cases}
\end{align}
Then, provided that $\cos\overline\theta=c(\overline u)$ and $\sin\overline\theta=s(\overline u)$ at the
prebounce Kasner regime, and $\cos\widetilde\theta=c(\widetilde u)$ and $\sin\widetilde\theta=s(\widetilde u)$
after the bounce, the transition rule [Eq. \eqref{transition_theta_general}] is largely simplified:
\begin{equation}
\label{trans_u_general}
\widetilde u=\begin{cases}
-\overline u, &\overline u>0,
\\
-\overline u-2, &\overline u<-1,
\\
-\dfrac{\overline u}{1+2\overline u},	&\overline u\in(-1,0).
\end{cases}
\end{equation}
This map is continuous, and its fixed points are
$-1, 0$, and $\infty$, which respectively correspond
to the values $0, 2\pi/3$, and $4\pi/3$ of the angle $\theta$
that define the exit points.
In addition, from this result, and taking into account  Eq. \eqref{sectors_u}, it is also straightforward to predict
in which sector the next bounce will take place in terms of the initial value of $\overline u$.
Finally, let us point out that this map has been extensively used to prove
the chaotic character of the Mixmaster model---see, e.g., Refs. \cite{BKL,Belinski,Cornish-levin}. In particular,
it is known that after a finite number of iterations, integer values of $u$
end up in one of the exit points; whereas the quadratic irrational numbers
follow periodic orbits and form the repeller set.

\subsection{A sequence of bounces against the potential walls}\label{subsec:sequence_bounces}

In order to end this section, let us briefly extend the previous analysis to a sequence of $n$ bounces.
The values of the Kasner parameters after a sequence of $n$ transitions
can be given in terms of the initial Kasner parameters
simply by applying the above map [Eq. \eqref{trans_IX_simp}] $n$ times. That is, provided that the
Kasner parameters after $n$ transitions are given by the array
$\left(c_+^{(n)},c_-^{(n)},p_+^{(n)},p_-^{(n)}\right)$,
they will be related to the initial Kasner parameters $\left(c_+^{(0)},c_-^{(0)},p_+^{(0)},p_-^{(0)}\right)$ by
the relation,
\begin{align}
\label{trans_n_IX}
\left(c_+^{(n)},c_-^{(n)},p_+^{(n)},p_-^{(n)}\right)=T^n
\left(c_+^{(0)},c_-^{(0)},p_+^{(0)},p_-^{(0)}\right)= \underbrace{T\circ\dots\circ T}_{n}\left(c_+^{(0)},c_-^{(0)},p_+^{(0)},p_-^{(0)}\right).
\end{align}
This equation can be recursively applied, but note in particular that
the map $T$ depends on the value of the angle $\theta$ in the corresponding
prebounce phase. That is, depending on the sector where the bounce happens, the transition
$T$ is defined by a rotation $R^0$, $R$, or $R^{-1}$.
However, it is possible to explicitly obtain the rotation that one has to apply at
the $n$th bounce
in terms of the initial angle $\theta^{(0)}$. For this purpose, the first step is to parametrize $\theta^{(0)}$ in terms of $u^{(0)}$ as in Eq. \eqref{parametrization_u}. Then, we map this parameter to the positive real line by means of
the following map $r$ and decompose it as a continued fraction:
\begin{equation}\label{u_decompos} 
r\left(u^{(0)}\right)=
m_0+\frac{1}{m_1+\frac{1}{m_2+\frac{1}{m_3+\dots}}},
\quad\text{where}\quad r(u):=\begin{cases}
u, &u>0,
\\
-1-\dfrac{1}{u}, &u\in(-1,0),
\\
-\dfrac{1}{u+1},&u<-1,
\\
\end{cases}
\end{equation}
$m_0=\lfloor r(u^{(0)})\rfloor$ is the integer part of $r(u^{(0)})$, and $m_i$ are nonvanishing integers. Then, we just need to apply iteratively the transition rule for $u$ [Eq. \eqref{trans_u_general}] and combine it with the relation \eqref{sectors_u}.
In this way, we obtain that the $n$th bounce will take place in the sector
\begin{align}\label{sector_n}
S_{\text{mod}(s_{n-1},3)+1},
\end{align}
where $\text{mod}$ is the modulo operation, $s_{n-1}:=\sum_{k=1}^{n-1}[2-\text{mod}(k-l_k,2)]+m$,
and $l_k$ is the non-negative integer that satisfies
$k\in(m_0+\dots+m_{{l_k}-1},m_0+\dots+m_{l_k}]$.
Therefore, the rotation matrix to be considered for the $n$th transition is
\begin{equation}
R^{\text{mod}(s_{n-1},3)}.
\end{equation}
In other words, the Kasner parameters after $n$ transitions are given in terms
of the initial Kasner parameters as,
\begin{align}\nonumber
\left(c_+^{(n)}\!,c_-^{(n)}\!,p_+^{(n)}\!,p_-^{(n)}\right)&=T^{n}
\left(c_+^{(0)}\!,c_-^{(0)}\!,p_+^{(0)}\!,p_-^{(0)}\right)
\\
&=R^{-\text{mod}(s_{n-1},3)}\!\circ B\circ R^{\text{mod}(s_{n-1},3)}
\!\circ\!\dots\!\circ R^{-\text{mod}(s_{0},3)}\!\circ B \circ R^{\text{mod}(s_{0},3)}
\!\left(c_+^{(0)}\!,c_-^{(0)}\!,p_+^{(0)}\!,p_-^{(0)}\right),
\nonumber
\end{align}
where the rotation matrices are now explicitly written in terms of the continued-fraction
decomposition of $u^{(0)}$ [Eq. \eqref{u_decompos}].

\section{Quantum analysis}\label{sec:quant}

Once we have provided the classical Kasner map [Eq. \eqref{trans_IX}] for our canonical variables,
in this section, we will analyze the quantum system in order to obtain the corresponding
Kasner map. The information of the quantum state
can be completely encoded in its moments, defined
as the expectation value of powers of the basic operators. In order to
describe our system, we will use the central moments:
\begin{equation}
\label{moments}
\Delta(\beta_+^i\,p_+^j\,\beta_-^k\,p_-^l):=\Big\langle
(\hat{\beta}_+-\beta_+)^i
(\hat{p}_+- {p}_+)^j
(\hat{\beta}_--\beta_-
)^k(\hat{p}_--{p}_-
)^l
\Big\rangle_{\text{Weyl}},
\end{equation}
where the Weyl subscript stands for a completely symmetric ordering of the operators;
$i$, $j$, $k$ and $l$ are non-negative integers; and we have defined the expectation
values $\beta_+:=\langle \hat{\beta}_+\rangle$, $p_+:=\langle \hat{p}_+\rangle$,
$\beta_-:=\langle \hat{\beta}_-\rangle$, and  $p_-:=\langle \hat{p}_-\rangle$.
The dimensions of such moments are given by $\hbar^{(i+j+k+l)/2}$, and thus
the sum of the integers $i+j+k+l$ will be referred to as the order of the corresponding moment.
Note that the first nontrivial order is the second one, as the zeroth-order moment would
just give the unit norm of the normalized wave function, whereas the first-order moments are trivially vanishing.

In this framework, the dynamical flow of states is described by the effective Hamiltonian,
defined as the expectation value of the Hamiltonian operator $\hat{H}$. One can implement
a formal power expansion around the expectation values, and obtain this effective Hamiltonian
in terms of quantum moments:
\begin{align}
H_Q&:=
\big\langle
\hat{H}(\hat{\beta}_+,\hat{p}_+,\hat{\beta}_-,\hat{p}_-)
\big\rangle 
=H+
\sum_{i+j+k+l=2}^{+\infty} \frac{1}{i!j!k!l!}
\dfrac{\partial^{i+j+k+l}H(\beta_+,p_+,\beta_-,p_-)}{\partial\beta_+^i\partial p_+^j\partial\beta_-^k\partial p_-^l}\Delta(\beta_+^ip_+^j\beta_-^kp_-^l),
\label{effective_hamiltonian}
\end{align}
where $H=H(\beta_+,p_+,\beta_-,p_-)$ is the classical Hamiltonian of the system [Eq. \eqref{hamiltonian}],
and the ordering of the Hamiltonian operator has been chosen to be completely symmetric.

In order to obtain the evolution equations for the different variables, one can just
compute the Poisson brackets with the above Hamiltonian. For the expectation values,
one then gets
\begin{align}
\label{quantum_em_beta_+}
\frac{d\beta_+}{d\alpha}
=\{
\beta_+,H_Q
\}&=\frac{\partial H_Q}{\partial p_+},
\\
\label{quantum_em_beta_-}
\frac{d\beta_-}{d\alpha}
=\{
\beta_-,H_Q
\}&=\frac{\partial H_Q}{\partial p_-},
\\
\label{quantum_em_p_+}
\frac{dp_+}{d\alpha}
=\{
p_+,H_Q
\}&=-\frac{\partial H_Q}{\partial\beta_+},
\\
\label{quantum_em_p_-}
\frac{dp_-}{d\alpha}
=\{
p_-,H_Q
\}&=-\frac{\partial H_Q}{\partial\beta_-},
\end{align}
whereas the time derivative of the moments is given by
\begin{align}
\frac{d\Delta(\beta_+^ip_+^j\beta_-^kp_-^l)}{d\alpha}
&=\{
\Delta(\beta_+^ip_+^j\beta_-^kp_-^l),H_Q
\}
\label{quantum_em_moments}
\\
\nonumber
&=
\sum_{m+n+r+s=2}^{+\infty} \frac{1}{m!n!r!s!}
\dfrac{\partial^{m+n+r+s}H(\beta_+,p_+,\beta_-,p_-)}{\partial\beta_+^m\partial p_+^n\partial\beta_-^r\partial p_-^s}\{\Delta(\beta_+^ip_+^j\beta_-^kp_-^l),\Delta(\beta_+^mp_+^n\beta_-^rp_-^s)\},
\end{align}
with the Poisson brackets between any two expectation values defined in terms of the commutator
in the usual way i.e., $\{\langle{\hat{f}}\rangle,\langle{\hat{g}}\rangle\}:=\frac{1}{i\hbar}\langle[{\hat{f},\hat{g}}]\rangle$ for any operators $\hat{f}$ and $\hat{g}$. 

Here we will not prove that the quantum dynamics of the Bianchi IX model can be asymptotically
described as a sequence of Kasner epochs with bounces against the potential walls,
since already at the classical level, the
proof of such a statement is quite involved. We just want to point
out that this is a nontrivial assumption, as the quantum dynamics might generically change this
behavior. However, the numerical resolution that will be presented in
Sec. \ref{subsec:dynamical_evolution} indicates that at the semiclassical level
the model still exhibits this pattern. In addition, one can also study the validity of
the Kasner approximation in terms of the conditions that must be obeyed by the potential
$U$ and its derivatives, which in the classical system has
led us to the conditions in Eqs. \eqref{kasner_cond_1}--\eqref{kasner_cond_3}. Even if for the quantum system
it is very complex to obtain the general expression for such conditions, the structure
of the terms that must be negligible is a combination of terms of the forms in Eqs.
\eqref{kasner_cond_1}--\eqref{kasner_cond_3} multiplied by certain quantum moments. Thus, as long as the
system follows a semiclassical dynamics and the quantum moments are relatively
small, it is reasonable to assume that, if requirements \eqref{kasner_cond_1}--\eqref{kasner_cond_3} are met,
the quantum system will also be following a Kasner dynamics.

Therefore, in order to obtain the quantum Kasner map, in this section we will assume
that asymptotically the quantum dynamics is accurately
described as a sequence of Kasner epochs
connected by transitions that can be modeled by the Bianchi II potential $V=e^{-8\beta_+}/6$
and appropriate rotations. For such a purpose, in Sec. \ref{subsec:Kasner_quantum}, we will first solve the
equations of motion during a generic Kasner epoch and define the parameters that characterize such regimes.
Next, in Sec. \ref{subsec:quantum_sectors}, by studying the leading exponential
term of the potential responsible at the end of the Kasner epoch,
we will define the quantum sectors $(S_1,S_2, S_3)$ in terms of the angle $\theta$,
or equivalently the ratio $p_-/p_+$.
As will be explained below, contrary to the classical case where there are only three
isolated exit points, in the quantum model there will generically be some
finite exit channels---finite intervals of $\theta$
for which none of the exponential terms
of the potential diverge as $\alpha\rightarrow-\infty$---and thus the system will follow an infinite
Kasner epoch until the singularity.
Then, in Sec. \ref{subsec:quantum_kasner_map}, we will present the quantum transition
law up to second order in moments, which will relate all the parameters of subsequent Kasner epochs.
Finally, in Sec. \ref{subsec:quantum_u_map}, the transition law for the parameter $u$
will be given, which encodes the information about the chaotic character of the system.

\subsection{Kasner epochs}\label{subsec:Kasner_quantum}

During the kinetically dominated Kasner regimes, the classical Hamiltonian takes the simple
form $H=P:=(p_+^2+p_-^2)^{1/2}$, and thus the quantum Hamiltonian \eqref{effective_hamiltonian} reduces to,
\begin{equation}
H_Q=P+
\sum_{i+j=2}^{+\infty} \frac{1}{i!j!}
\dfrac{\partial^{i+j}P(p_+,p_-)}{\partial p_+^i\partial p_-^j}\Delta(p_+^ip_-^j).
\end{equation}
Since this Hamiltonian does not depend on the shape parameters,
during this period the expectation values of the momenta $p_+$ and $p_-$,
as well as all the moments unrelated to the shape parameters---that is, $\Delta (p_+^ip_-^j)$ for any $i+j\geq 2$---are constants of motion. Furthermore, these constant moments are the only ones that appear in the evolution equations for the expectation values of the shape parameters,
which read
\begin{align}
\frac{d\beta_{+}}{d\alpha}&=\sum_{i,j=0}^{+\infty}
\frac{1}{i!j!}\dfrac{\partial^{i+j+1}P}{\partial p_+^{i+1}\partial p_-^{j}}\Delta(p_+^ip_-^j),\\
\frac{d\beta_{-}}{d\alpha}&=\sum_{i,j=0}^{+\infty}
\frac{1}{i!j!}\dfrac{\partial^{i+j+1}P}{\partial p_+^{i}\partial p_-^{j+1}}\Delta(p_+^ip_-^j).
\end{align}
Therefore, one can immediately integrate these equations and obtain
a linear evolution in time for the shape parameters,
\begin{align}
\label{quantum_kasner_beta_+_full}
&\beta_{+}=\left(\frac{p_+}{P}+\sum_{i+j=2}^{+\infty}
\frac{1}{i!j!}\dfrac{\partial^{i+j+1}P}{\partial p_+^{i+1}\partial p_-^{j}}\Delta(p_+^ip_-^j)\right)\alpha+c_+,
\\
\label{quantum_kasner_beta_-_full}
&\beta_{-}=\left(\frac{p_-}{P}+\sum_{i+j=2}^{+\infty}
\frac{1}{i!j!}\dfrac{\partial^{i+j+1}P}{\partial p_+^{i}\partial p_-^{j+1}}\Delta(p_+^ip_-^j)\right)\alpha+c_-,
\end{align}
with integration constants $c_+$ and $c_-$. As compared to the classical evolution in Eqs.
\eqref{beta_+_kas} and \eqref{beta_-_kas}, the system still follows a straight-line trajectory
in the $(\beta_+,\beta_-)$ plane, but with the slope modified by the fluctuations
of the momenta $p_+$ and $p_-$ and their correlations.

In summary, we have exactly obtained the quantum evolution of the expectation values
$\beta_+$, $\beta_-$, $p_+$, $p_-$, and of the moments unrelated to the shape parameters.
However, the evolution equations for the rest of the moments are very intricate,
as they form an infinite coupled system. Thus, in order to perform the analysis
of the Kasner quantum map, we will, from now on, assume a peaked semiclassical state,
for which third- and higher-order moments are negligible during the whole considered evolution. That is, we will
introduce a truncation and neglect any moment
$\Delta(\beta_+^i p_+^j \beta_-^k p_-^l)$ with $i+j+k+l\geq 3$, which corresponds to
dropping terms of order $\hbar^{3/2}$ and higher-order powers of $\hbar$. Under this assumption,
the slope of the straight line followed by the system in the $(\beta_+,\beta_-)$
plane reads
\begin{equation}
\tan\theta+\left(\frac{\Delta(p_+^2)}{P^2}-\frac{\Delta(p_-^2)}{P^2} \right)\tan\theta
-\frac{\Delta(p_+p_-)}{P^2}(1-\tan^2\theta),
\end{equation}
with the angle $\theta$ defined as in the classical case, $\cos\theta:=-\frac{p_+}{P}$ and $\sin\theta:=-\frac{p_-}{P}$.
Moreover, it is straightforward to solve the equations of motion for all the second-order moments.
In particular, the evolution of a moment of the form $\Delta(\beta_+^i p_+^j \beta_-^k p_-^l)$
will be given by a polynomial of order $i+k$ in $\alpha$. Therefore, the four correlations between
shape parameters $(\beta_+,\beta_-)$ and momenta $(p_+,p_-)$ are linear functions in $\alpha$,
\begin{align}
\label{quantum_kasner_delta_beta_+_p_+}
&\Delta(\beta_+p_+)=
\left(
\frac{\partial^{2}P}{\partial p_+^{2} }\Delta(p_+^{2})+
\frac{\partial^{2}P}{\partial p_+\partial p_- }\Delta(p_+p_-)
\right)\alpha+k_{\scriptsize_{++}}, 
\\
\label{quantum_kasner_delta_beta_+_p_-}
&\Delta(\beta_+p_-)=
\left(
\frac{\partial^{2}P}{\partial p_+^{2} }\Delta(p_+p_-)+
\frac{\partial^{2}P}{\partial p_+\partial p_- }\Delta(p_-^2)
\right)\alpha+k_{\scriptsize_{+-}},
\\
\label{quantum_kasner_delta_beta_-_p_+}
&\Delta(p_+\beta_-)=
\left(
\frac{\partial^{2}P}{\partial p_-^{2} }\Delta(p_+p_-)+
\frac{\partial^{2}P}{\partial p_+\partial p_- }\Delta(p_+^2)
\right)\alpha+k_{\scriptsize_{-+}},
\\
\label{quantum_kasner_delta_beta_-_p_-}
&\Delta(\beta_-p_-)=
\left(
\frac{\partial^{2}P}{\partial p_-^{2} }\Delta(p_-^{2})+
\frac{\partial^{2}P}{\partial p_+\partial p_- }\Delta(p_+p_-)
\right)\alpha+k_{\scriptsize_{--}}, 
\end{align}
whereas the fluctuations of the shape parameters and their correlation $\Delta(\beta_+\beta_-)$
are second-order polynomials in $\alpha$:
\begin{align}
\nonumber
\Delta(\beta_+^2)&=\left[
\left(\dfrac{\partial^{2}P}{\partial p_+^2}\right)^2
\Delta(p_+^2)+
\left(\dfrac{\partial^{2}P}{\partial p_+\partial p_-}\right)^2
\Delta(p_-^2)+
2\left(
\dfrac{\partial^{2}P}{\partial p_+^{2}}\right)
\left(
\dfrac{\partial^{2}P}{\partial p_+\partial p_-}
\right)
\Delta(p_+p_-)
\right]\alpha^2
\\
&
+2\left(
\dfrac{\partial^{2}P}{\partial p_+^{2}}k_{\scriptsize_{++}}+
\dfrac{\partial^{2}P}{\partial p_+\partial p_-}k_{\scriptsize_{+-}}\right)\alpha 
+c_{\scriptsize_{++}},
\label{quantum_kasner_delta_beta_+_2}
\\
\nonumber
\Delta(\beta_-^2)&=
\left[
\left(\dfrac{\partial^{2}P}{\partial p_-^2}\right)^2
\Delta(p_-^2)+
\left(\dfrac{\partial^{2}P}{\partial p_+\partial p_-}\right)^2
\Delta(p_+^2)+
2\left(
\dfrac{\partial^{2}P}{\partial p_-^{2}}\right)
\left(
\dfrac{\partial^{2}P}{\partial p_+\partial p_-}
\right)
\Delta(p_+p_-)
\right]\alpha^2
\\
&+2\left(
\dfrac{\partial^{2}P}{\partial p_-^{2}}k_{\scriptsize_{--}}+
\dfrac{\partial^{2}P}{\partial p_+\partial p_-}k_{\scriptsize_{-+}}\right)\alpha+c_{\scriptsize_{--}},
\label{quantum_kasner_delta_beta_-_2}
\\
\nonumber
\Delta(\beta_+\beta_-)&=\left[
\dfrac{\partial^{2}P}{\partial p_-^{2}}
\dfrac{\partial^{2}P}{\partial p_+\partial p_-}
\Delta(p_-^2)+
\dfrac{\partial^{2}P}{\partial p_+^{2}}
\dfrac{\partial^{2}P}{\partial p_+\partial p_-}
\Delta(p_+^2)
+\left[
\left(\dfrac{\partial^{2}P}{\partial p_+\partial p_-}\right)^2
+\dfrac{\partial^{2}P}{\partial p_+^{2}}\dfrac{\partial^{2}P}{\partial p_-^{2}}
\right]\Delta(p_+p_-)
\right]\alpha^2
\\
&
+\left[
\dfrac{\partial^{2}P}{\partial p_-^{2}}k_{\scriptsize_{+-}}+
\dfrac{\partial^{2}P}{\partial p_+^{2}}k_{\scriptsize_{-+}}
+
\dfrac{\partial^{2}P}{\partial p_+\partial p_-}\left(k_{\scriptsize_{++}}+k_{\scriptsize_{--}}\right)
\right]\alpha 
+c_{\scriptsize_{+-}}.
\label{quantum_kasner_delta_beta_+_beta_-}
\end{align}
Note that, in order to write down these solutions we have defined the integration constants
$k_{\scriptsize_{\pm\pm}}$ for the correlations between momenta and shape parameters [Eqs. \eqref{quantum_kasner_delta_beta_+_p_+}--\eqref{quantum_kasner_delta_beta_-_p_-}],
and $c_{\scriptsize_{\pm\pm}}$ for the fluctuations and the correlation between the shape parameters
[Eqs. \eqref{quantum_kasner_delta_beta_+_2}--\eqref{quantum_kasner_delta_beta_+_beta_-}].
Since at some points it will be convenient to refer to all these constants, along with
$\Delta(p_+^ip_-^j)$, collectively, let us define the Kasner parameters $K_{ijkl}$:
\begin{align}
\label{constans_K_def}
\begin{split}
	& K_{0200}:=\Delta(p_+^2),\quad
	K_{0002}:=\Delta(p_-^2),\quad
	K_{0101}:=\Delta(p_+p_-),
	\\
	&K_{1100}:=k_{\scriptsize_{++}}, \quad
	K_{1001}:=k_{\scriptsize_{+-}}, \quad
	K_{0110}:=k_{\scriptsize_{-+}},\quad
	K_{0011}:=k_{\scriptsize_{--}},
	\\
	&K_{2000}:=c_{\scriptsize_{++}}, \quad
	K_{0020}:=c_{\scriptsize_{\scriptsize_{--}}},\quad
	K_{1010}:=c_{\scriptsize_{+-}}.
\end{split}
\end{align}
That is, $K_{ijkl}$ is just the value the moment $\Delta(\beta_+^ip_+^j\beta_-^kp_-^l)$
would take at $\alpha=0$ by extrapolating its corresponding Kasner evolution. (Note that
we are assuming that
$\alpha\rightarrow-\infty$, and thus $\alpha$ will be different from zero during
a typical Kasner epoch.) Therefore, up to second order in moments,
the parameters that characterize each quantum Kasner epoch will be the four constants
$c_+$, $c_-$, $p_+$, and $p_-$,
as in the classical case, along with the 10 constants $K_{ijkl}$
with $i+j+k+l=2$, which encode the dynamical information of the moments.

\subsection{Quantum sectors $S_1$, $S_2$, and $S_3$}\label{subsec:quantum_sectors}

\begin{figure}[t]
	\centering
	\includegraphics[width=0.7\linewidth]{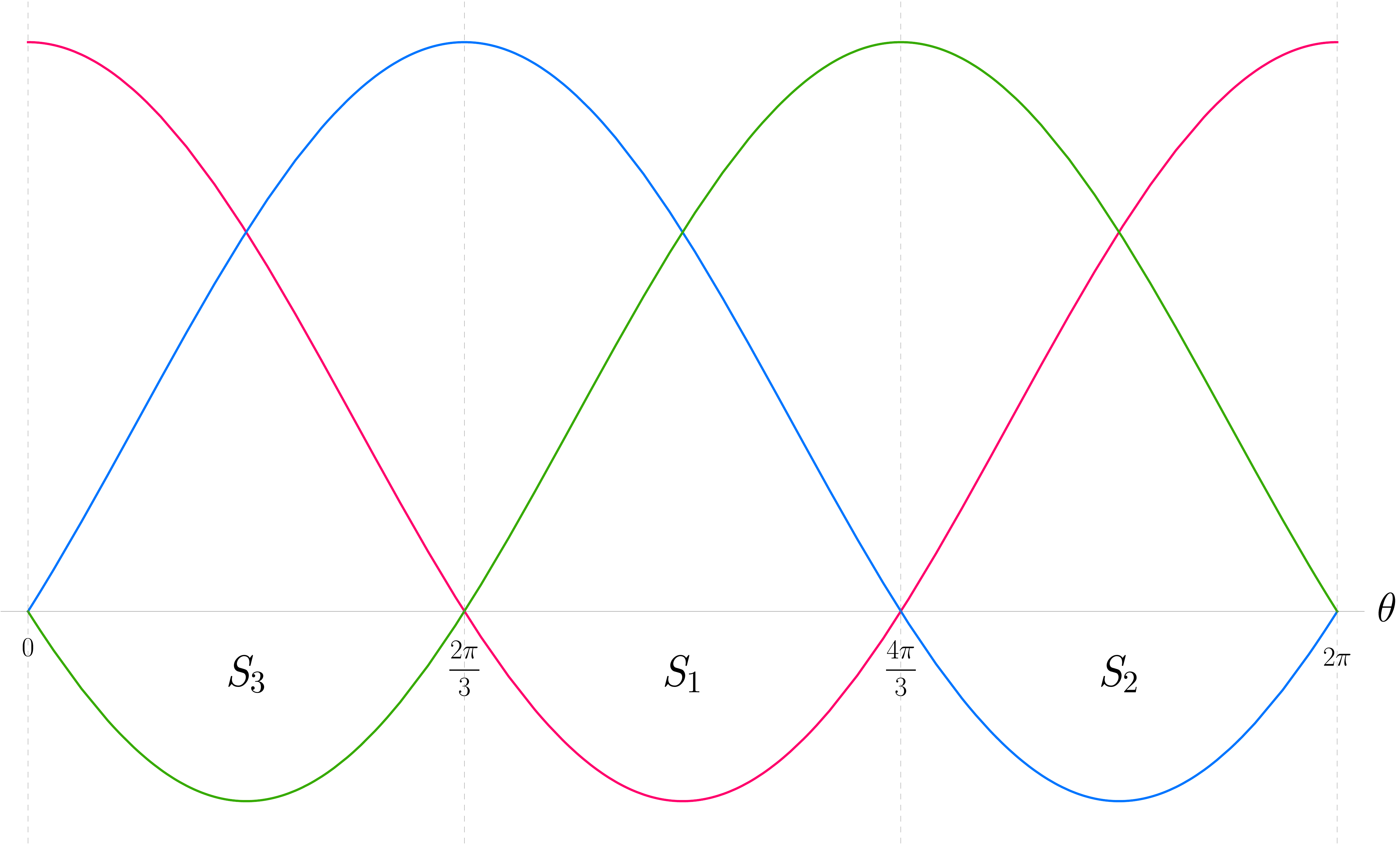}
	\caption{ In this plot, we show the
		coefficients multiplying $\alpha$ on the  exponents
		of the three leading exponential terms of the potential
		($e^{4\alpha-8\beta_+}$, $e^{4\alpha-4 \sqrt{3} \beta_-+4 \beta_+}$, and $e^{4\alpha+4 \sqrt{3} \beta_-+4 \beta_+}$) during a classical Kasner regime [Eqs. \eqref{beta_+_kas_theta} and \eqref{beta_-_kas_theta}] in terms of the angle $\theta$.
		More explicitly, these coefficients are, respectively,  $4+8\cos\theta$ (in pink), $4+4\sqrt{3}\sin\theta-4\cos\theta$ (in blue), and $4-4\sqrt{3}\sin\theta-4\cos\theta$ (in green). Since we are studying the evolution of the system toward $\alpha\to -\infty$, a given exponential term will diverge when its corresponding coefficient is negative. This condition
		defines the three classical sectors $S_1$, $S_2$, and $S_3$ as the ranges of $\theta$  $(2\pi/3,4\pi/3)$,$(4\pi/3,2\pi)$, and $(0,2\pi/3)$, respectively. At the boundary points $\theta= 0$, $2\pi/3$, and $4\pi/3$, all the exponents are convergent, and thus the system
		will follow an infinite Kasner regime.}
	\label{fig:exponential_ranges_classic}
\end{figure}

Generically, the above Kasner dynamics will come to an end as soon as the potential increases and produces a bounce. As we have explained in the
previous section, the prebounce angle $\theta$, defined in terms of the ratio $p_-/p_+$,
completely determines in which sector, $S_1$, $S_2$, or $S_3$, the bounce will happen, or equivalently,
which exponential term will produce it: $e^{4\alpha-8\beta_+}$, $e^{4\alpha-4 \sqrt{3} \beta_-+4 \beta_+}$, or $e^{4\alpha+4 \sqrt{3} \beta_-+4 \beta_+}$, respectively. In the classical case, taking into account the Kasner dynamics [Eqs. \eqref{beta_+_kas_theta} and \eqref{beta_-_kas_theta}], for a given value of $\theta$,
one of these exponential terms is diverging, whereas the other two tend to zero as $\alpha\rightarrow -\infty$ (see Fig. \ref{fig:exponential_ranges_classic}). 
The only exceptions are the values $\theta= 0, 2\pi/3$, and $4\pi/3$, for which none of the three exponential terms
diverges as $\alpha\rightarrow -\infty$ (one of them tends to zero, while the other two converge to some finite value).
For these exit points, there is no bounce, and the system follows a Kasner dynamics until the singularity. 
In the quantum scenario, however, the situation is more involved, and, as will be explained below, depending
on the values of the quantum moments, there will be finite ranges of $\theta$ for which all three exponential terms tend to zero, or other ranges for which two of the exponential
terms diverge and the third one tends to zero. The former will describe a finite open exit channel, whereas
the latter will imply that the corresponding exit channel is closed.

Let us be more specific. Following the classical analysis,
we define the angle $\theta\in[0, 2\pi)$ as $\cos\theta:=-p_+/P$ and $\sin\theta:=-p_-/P$,
and taking into account the dynamics given by Eqs. \eqref{quantum_kasner_beta_+_full} and \eqref{quantum_kasner_beta_-_full},
we will define the quantum sectors $S_1$, $S_2$, and $S_3$ as the ranges of values of $\theta$ for which
the corresponding exponential term---i.e.,
$e^{4\alpha-8\beta_+}$, $e^{4\alpha-4 \sqrt{3} \beta_-+4 \beta_+}$, or $e^{4\alpha+4 \sqrt{3} \beta_-+4 \beta_+}$---diverges faster than the other two as $\alpha\rightarrow -\infty$. The inequalities obtained by this condition
involve expressions with trigonometric functions of $\theta$, which are not immediate to solve. However, since
in our semiclassical analysis a small value of the moments is assumed, one can
linearize the problem around the corresponding classical solution, which leads to
the following ranges of $\theta$ for each of the quantum sectors:
\begin{align}
\label{quant_sectors}
\begin{split}
S_1:\quad
&\theta_{1}^{\min}:=\frac{2\pi}{3}+\max
\left\{
3\vp-
5\vm
-\corr
,2\vp
-6\vm
-2\corr
\right\}
\\
&\hspace{1.2cm}
< \theta
<\frac{4\pi}{3}+
\min
\left\{
-3\vp+
5\vm
-r
,-2\vp
+6\vm
-2\corr
\right\}=:
\theta_{1}^{\max},
\\
S_2: \quad
&\theta_{2}^{\min}:=\frac{4\pi}{3}+\max
\left\{
-\vp+
7\vm
-3\corr
,-2\vp
+6\vm
-2\corr
\right\}
< \theta <
2\pi- 4(\vm-\corr)=:\theta_{2}^{\max},
\\
S_3: \quad
&
\theta_{3}^{\min}:=
4(\vm+\corr)
<\theta<\frac{2\pi}{3}+\min
\left\{
\vp-
7\vm
-3\corr
,2\vp
-6\vm
-2\corr
\right\}=:
\theta_{3}^{\max}
,
\end{split}
\end{align}
where, for compactness, we have defined the relative moments $\vp:=\frac{\sqrt{3}}{8}\frac{\Delta(p_+^2)}{P^2}$, $\vm:=\frac{1}{8\sqrt{3}}\frac{\Delta(p_-^2)}{P^2}$, and $r:=\frac{1}{4}\frac{\Delta(p_+p_-)}{P^2}$.
The maximum and minimum functions that appear in some of the limits come from the fact that the
boundaries of a sector are given as the largest and smallest
angles for which the corresponding exponential term either converges to a nonzero value or diverges as
fast as the exponential term of the adjacent sector.
In particular, in the above maximum (minimum) functions, the first argument is the smallest
(largest) angle for which the exponential term in question converges to a nonzero value,
while the second argument is
the smallest
(largest) angle for which it evolves at the same rate as the exponential of the adjacent sector.
Note that for the definitions of $\theta_2^{\rm max}$ (the upper limit of $S_2$) and of
$\theta_3^{\rm min}$ (the lower limit of $S_3$), the use of maximum and minimum functions
could be avoided due to the positive definiteness of $v_{-}$. More precisely, $\theta_2^{\rm max}$
is obtained to be $\theta_2^{\rm max}=2\pi+\min\{4 (r-v_-),4r\}=2\pi-4(v_-r)$, whereas
$\theta_3^{\rm min}=\max\{4 (v_-+r),4r\}=4(v_-+r)$.

\begin{figure}[t]
	\centering
	\includegraphics[width=0.8\linewidth]{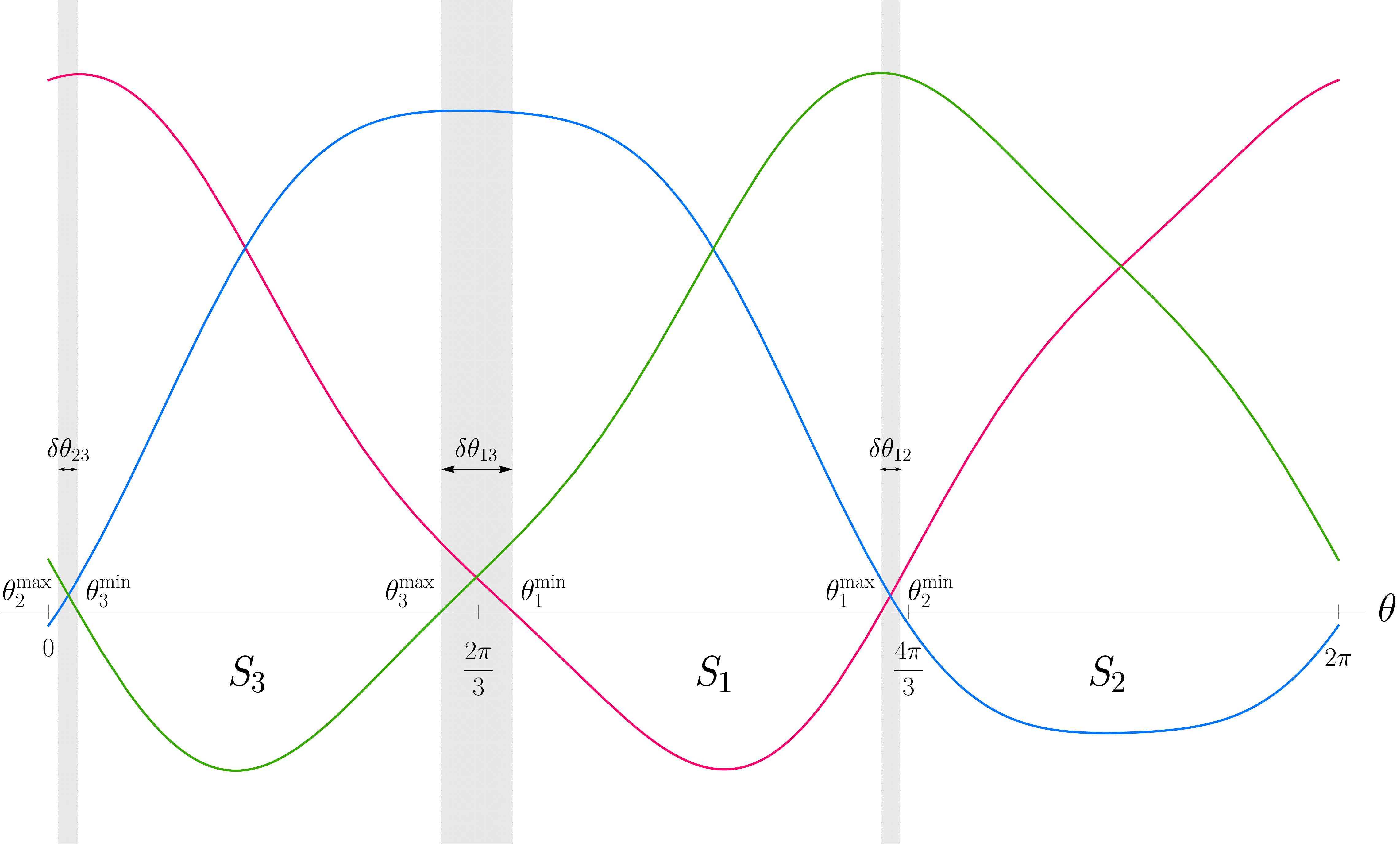}
	\caption{In this plot, we show the
			coefficients multiplying $\alpha$ on the exponents
			of the three leading exponential terms of the potential:
			$e^{4\alpha-8\beta_+}$ (in pink), $e^{4\alpha-4 \sqrt{3} \beta_-+4 \beta_+}$ (in blue), and $e^{4\alpha+4 \sqrt{3} \beta_-+4 \beta_+}$ (in green) during a quantum Kasner regime [Eqs. \eqref{quantum_kasner_beta_+_full} and \eqref{quantum_kasner_beta_-_full}] in terms of the angle $\theta$. A second-order truncation of the moments is considered, with numerical values $v_+=\frac{3\sqrt{3}}{80}$,
			$v_-=\frac{1}{40\sqrt{3}}$, and $r=\frac{1}{40}$.
			Since we are studying the evolution of the system toward $\alpha\to -\infty$, a given exponential term will diverge when its corresponding coefficient is negative, and thus, this condition defines the three quantum sectors $S_1$, $S_2$, and $S_3$.}
	\label{fig:exponential_ranges_open}
\end{figure}
\begin{figure}[t]
	\centering
	\centering
	\begin{subfigure}[h]{0.327\textwidth}
		\centering
		\includegraphics[width=\textwidth]{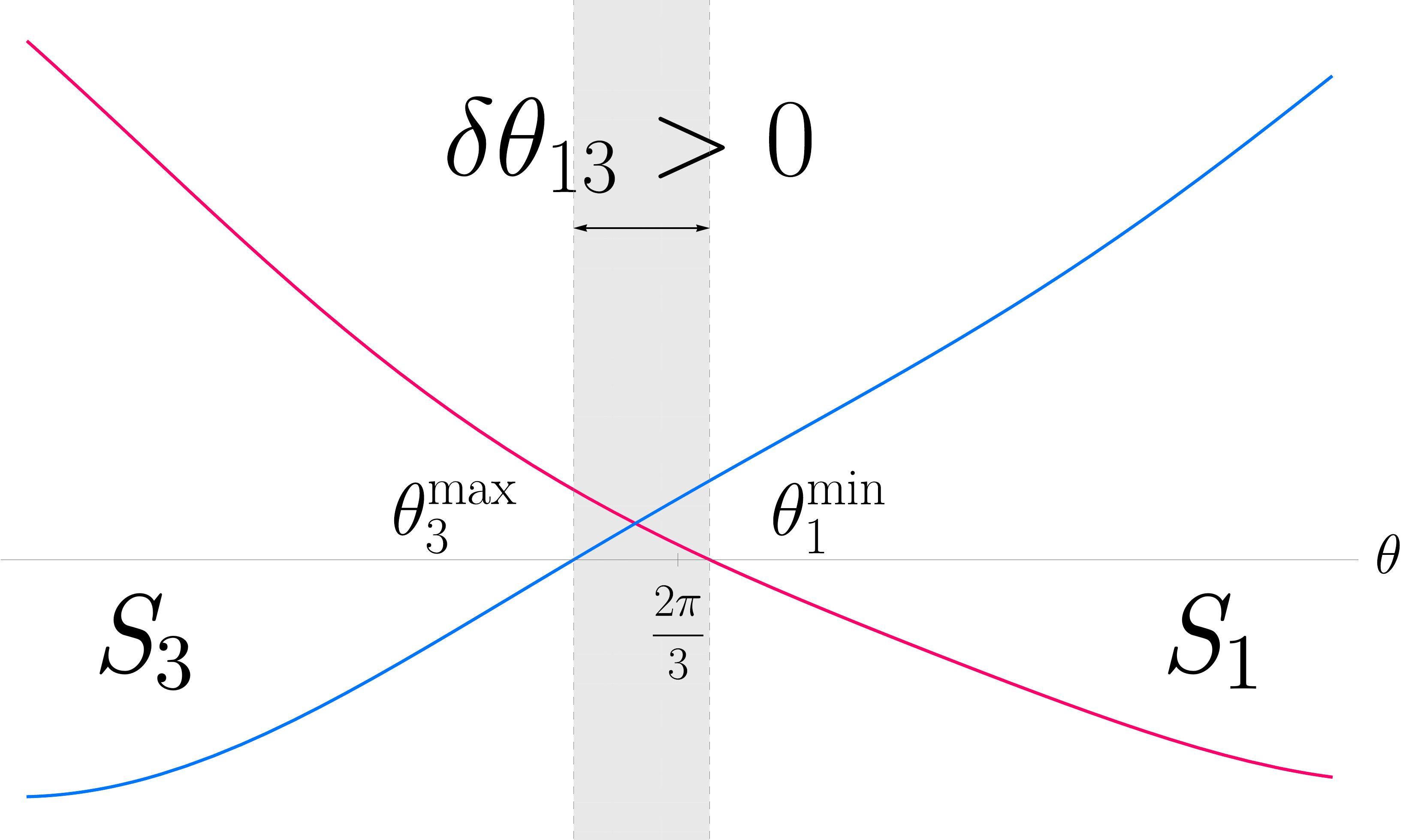}
	\end{subfigure}
	\hfill
	\begin{subfigure}[h]{0.327\textwidth}
		\centering
		\includegraphics[width=\textwidth]{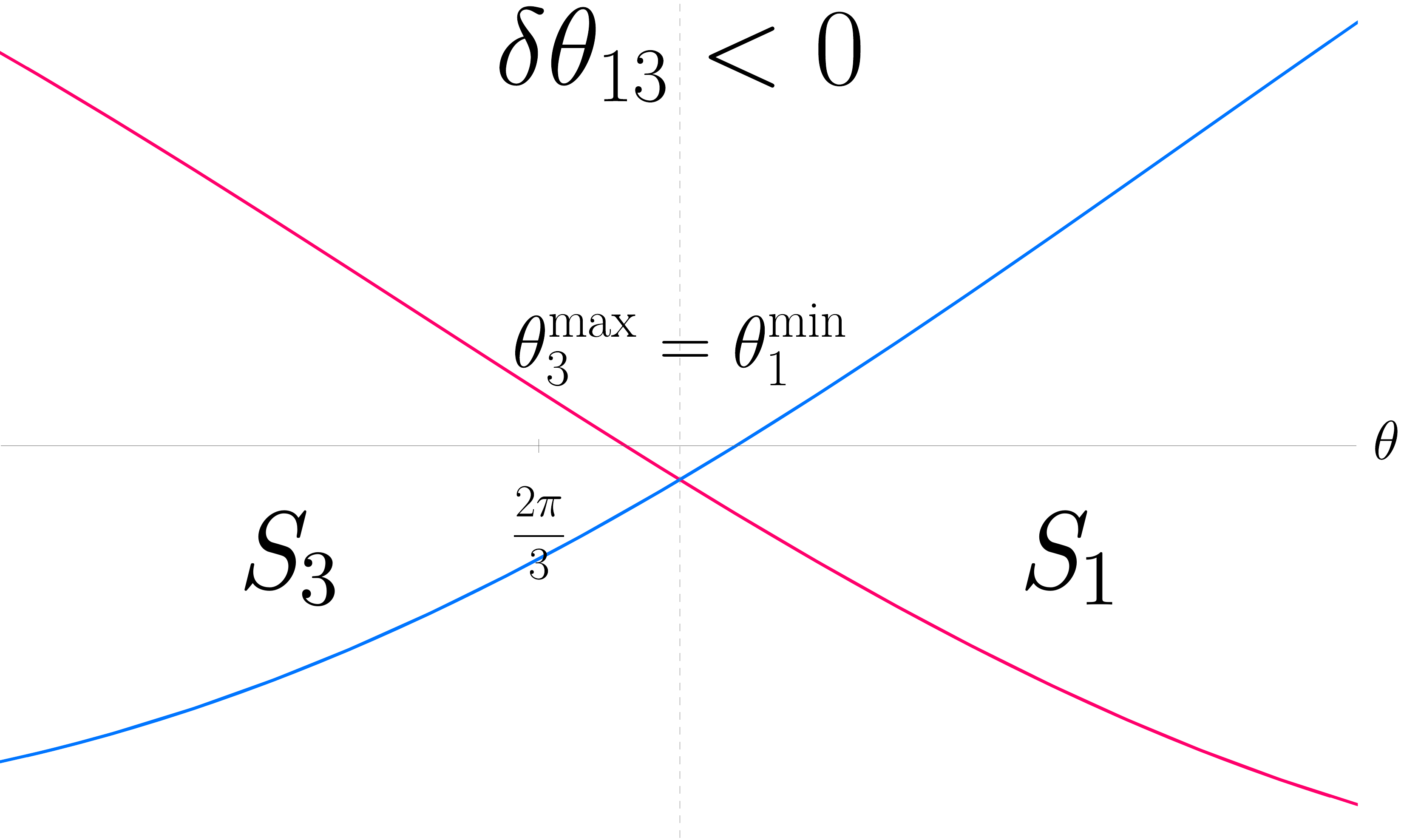}
	\end{subfigure}
	\begin{subfigure}[h]{0.327\textwidth}
		\centering
		\includegraphics[width=\textwidth]{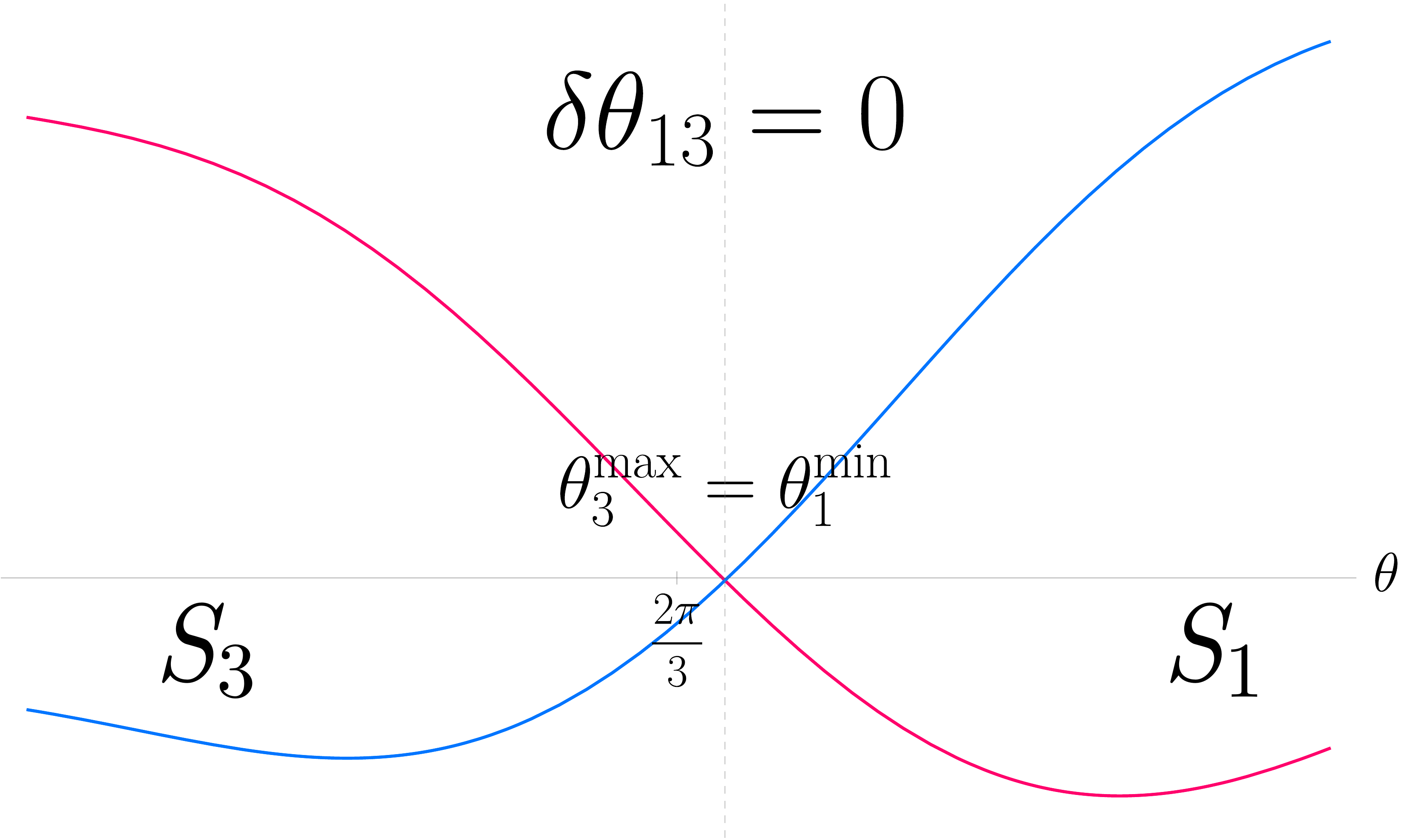}
	\end{subfigure}
	\caption{ In this plot, we show the
			coefficients multiplying $\alpha$ on the exponents
			of the three leading exponential terms related to $S_1$ and $S_3$: $e^{4\alpha-8\beta_+}$ (in pink) and $e^{4\alpha+4 \sqrt{3} \beta_-+4 \beta_+}$ (in blue), respectively, during a quantum Kasner regime [Eqs. \eqref{quantum_kasner_beta_+_full} and \eqref{quantum_kasner_beta_-_full}] in terms of the angle $\theta$. For all three plots, a second-order truncation of the moments has been considered. The first plot corresponds to  $v_+=\frac{\sqrt{3}}{40}$, $v_-=\frac{1}{40\sqrt{3}}$, and  $r=\frac{1}{40}$, and it shows an open exit channel
			with $\delta{\theta}_{13}>0$. The second one corresponds to $v_+=\frac{1}{40\sqrt{3}}$,  $v_-=\frac{1}{120\sqrt{3}}$, and $r=-\frac{1}{16}$, and it presents a fully closed channel with $\delta\theta_{13}<0$. Finally, 
			the third plot corresponds to $v_+=\frac{\sqrt{3}}{40}$,  $v_-=\frac{1}{40\sqrt{3}}$, and $r=-\frac{1}{10\sqrt{3}}$, which results in the limit case $\delta\theta_{13}=0$.}
	\label{fig:exponential_ranges_all}
\end{figure}

As anticipated above, there are finite ranges of $\theta$, namely
$(\theta_1^{\rm max},\theta_2^{\rm min})\cup(\theta_2^{\rm max},\theta_3^{\rm min})\cup (\theta_3^{\rm max},\theta_1^{\rm min})$, for which none of the exponential terms
diverge, and thus they do not correspond to any of the sectors, as shown in Fig. \ref{fig:exponential_ranges_open}
for a particular example. These values will form the exit channels
for which there will be no bounce, and the system will follow a Kasner dynamics
until the singularity. The angular width of each of these exit channels is given as follows:
\begin{align}\label{delta_escaping_channels_13}
&\delta\theta_{13}=
2\left(\vp+
\vm
+\corr\right),
\\
\label{delta_escaping_channels_12}
&\delta\theta_{12}=
2\left(\vp+
\vm
-\corr\right),
\\
\label{delta_escaping_channels_23}
&\delta\theta_{23}=8\vm,
\end{align}
where $\delta\theta_{ij}$ corresponds to the channel between the sectors $S_i$ and $S_j$.
The signs of these angular widths will determine whether the exit channels are open or closed
(see Fig. \ref{fig:exponential_ranges_all}).
On the one hand, if $\delta\theta_{ij}$ is positive, the corresponding exit channel will be
open, since all the exponential terms converge as $\alpha\rightarrow -\infty$, and there will be no bounce
of the system for any value of $\theta$ in this region.
On the other hand, if $\delta\theta_{ij}$ is negative, the exit channel
is closed, as the exponential terms of the adjacent sectors diverge at the same time; thus, for
any value of $\theta$ around this boundary 
the Kasner dynamics of the system [Eqs. \eqref{quantum_kasner_beta_+_full}--\eqref{quantum_kasner_delta_beta_+_beta_-}] will end up by an interaction with an exponential wall.
The limit case of a vanishing $\delta\theta_{ij}$ implies that the two exponential terms
of the adjacent sectors simultaneously converge to certain nonzero finite values as $\alpha\rightarrow-\infty$.
In the classical case, this leads to an exit point, since there are no other terms in the Hamiltonian
that might diverge. However, in the quantum effective Hamiltonian [Eq. \eqref{effective_hamiltonian}] there are fluctuations
and correlations of the shape parameters $\Delta(\beta^np^m)$ that diverge as $\alpha^n$; see Eqs. \eqref{quantum_kasner_delta_beta_+_p_+}--\eqref{quantum_kasner_delta_beta_+_beta_-}.
In principle, the dynamics of these limit points cannot be approximated by an exponential Bianchi II potential,
and they are very sensitive to the truncation order, since a higher-order truncation would lead
to new divergent terms. Hence, a specific higher-order analysis would be needed to clearly understand these particular
cases. Nonetheless, as they are a very particular set of cases, their relevance in the overall properties of
the model is expected to be very limited, and we will exclude them from the subsequent analysis.

In particular, it is interesting to note that the channel between $S_2$ and $S_3$ is always open,
which is clearly seen in Eq. \eqref{delta_escaping_channels_23}, as the fluctuation $v_{-}$ is positive definite.
This is due to the fact that the exponential terms corresponding to these two sectors cannot be simultaneously divergent
as $\alpha\rightarrow-\infty$.
However, one of the other two channels will be closed if the correlation is relatively large and obeys
$|r|\geq v_++v_-$.

Since the fluctuations $v_+$ and $v_-$ are positive,
they both contribute to widen the exit channels [Eqs. \eqref{delta_escaping_channels_13}--\eqref{delta_escaping_channels_23}].
On the contrary, the correlation $r$ does not have a definite
sign, so it can either widen or narrow these exits. Nonetheless, the total angular width of the three
channels $\delta\theta_{12}+\delta\theta_{23}+\delta\theta_{31}$ is independent of the correlation, and it can be used
to define an exit probability. That is,
in the classical model, the probability of escaping the bouncing behavior
after any given bounce is zero, since one needs to take one of three fine-tuned values:
$\theta=0$, $2\pi/3$, or $4\pi/3$. However, for the quantum model, this probability
is nonzero, and it reads
\begin{align}\label{prob_escaping}
\frac{\delta\theta_{13}+\delta\theta_{12}+\delta\theta_{23}}{2\pi}=\frac{2}{\pi}(v_++3 v_-).
\end{align}

The presence of these finite exit channels is a key feature of the quantum model,
which introduces a qualitative difference from the classical one. In particular,
it shows how the quantum effects change the chaotic behavior of the Mixmaster model:
contrary to the classical case, given a set of points as initial data in an exit channel,
all of them will follow a qualitatively similar dynamics and their trajectories
in the phase space will not exponentially diverge. Therefore, at this scale,
the classical chaotic behavior will cease.

\begin{figure}[t]
	\centering
	\includegraphics[width=0.75\linewidth]{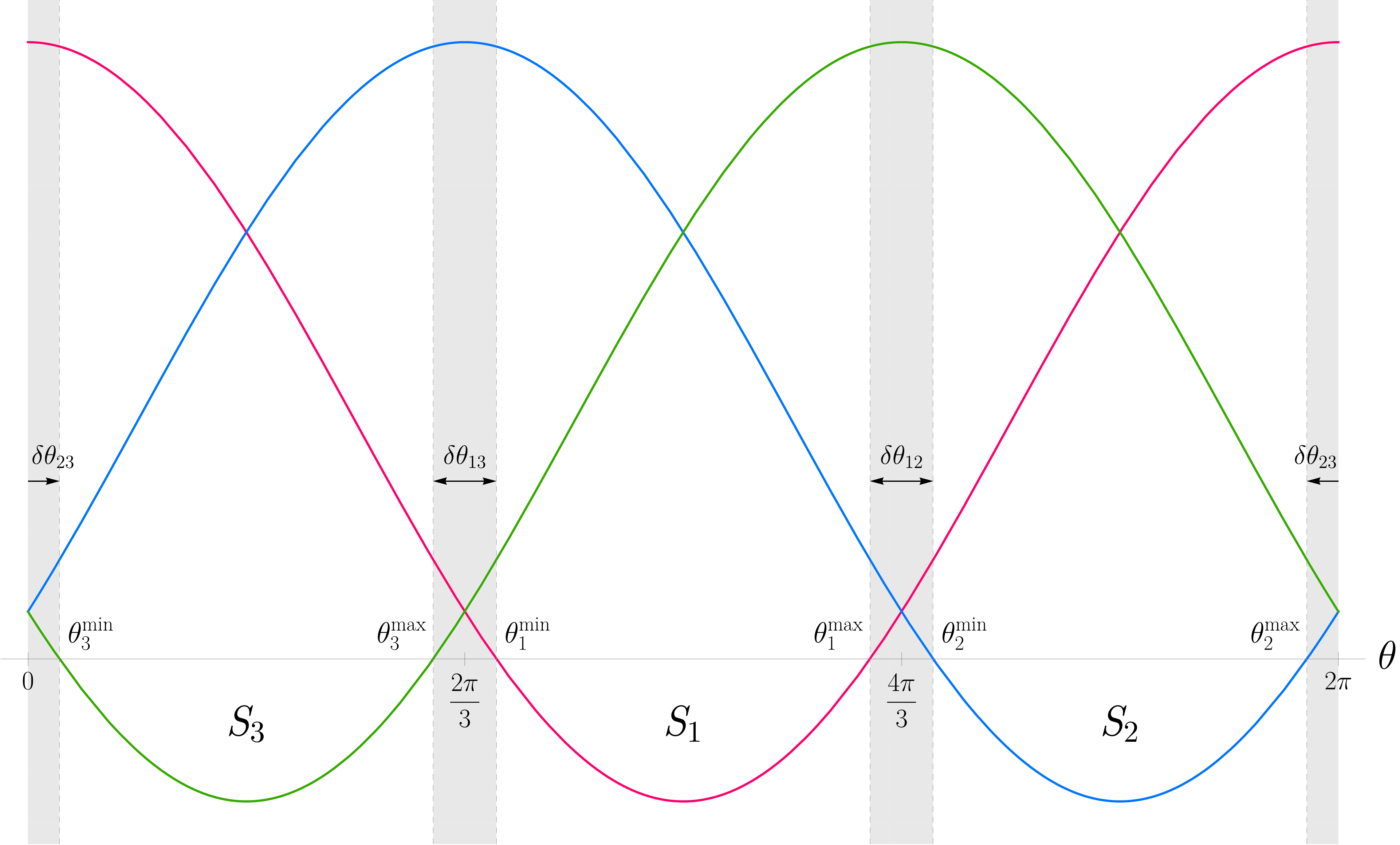}
	\caption
	{In this plot, we show the
		coefficients multiplying $\alpha$ on the  exponents
		of the three leading exponential terms of the potential:
		$e^{4\alpha-8\beta_+}$ (in pink), $e^{4\alpha-4 \sqrt{3} \beta_-+4 \beta_+}$ (in blue), and $e^{4\alpha+4 \sqrt{3} \beta_-+4 \beta_+}$ (in green), during a quantum Kasner regime [Eqs. \eqref{quantum_kasner_beta_+_full}-\eqref{quantum_kasner_beta_-_full}] in terms of the angle $\theta$. A second-order truncation of the moments is considered with the
		numerical values $v_+=3v_-=\frac{\sqrt{3}}{20}$ and $r=0$, which correspond to a symmetric state.
		}
	\label{fig:exponential_ranges_sym}
\end{figure}

Additionally, it is interesting to remark that the classification in Eq. \eqref{quant_sectors}---and all the properties derived therefrom---does not have the usual $2\pi n/3$ rotational symmetry of the system, even if the effective Hamiltonian \eqref{effective_hamiltonian} does. This is due to the fact that generically the quantum state (wave function) does not need
to be symmetric. The particular case of a symmetric wave function implies a vanishing correlation $r=0$
and the same fluctuation for both momenta $\Delta(p_+^2)=\Delta(p_-^2)$, which leads to $v_+=3 v_-$. In this case, the three exit channels are open,
and their width is given by $\delta\theta_{12}=\delta\theta_{13}=\delta\theta_{23}= 8v_-$. An example of
such a completely symmetric case is shown in Fig. \ref{fig:exponential_ranges_sym}.

Finally, making use of the relation \eqref{utheta} and neglecting second- and higher-order
powers of the relative moments $(\vp,\vm,\corr)$,
it is immediately possible to obtain the explicit values of the parameter $u$ that will define the boundaries
between the different sectors:
\begin{align}
\label{quant_sectors_u_bounds}
\begin{split}
&u_{1}^{\min}:=\frac{1}{\sqrt{3}}\max
\left\{
3\vp-5\vm-\corr
,
2\vp -6\vm -2\corr
\right\},
\\
&
u_{1}^{\max}:=-\frac{1}{2}
+\frac{1}{4\sqrt{3}}\min
\left\{-3\vp+5\vm-\corr
+\frac{12}{3\vp-5\vm+\corr},
-2\vp+6\vm-2\corr+
\frac{6}{\vp-3\vm+\corr}
\right\}
,
\\
& u_{2}^{\min}:=
-\frac{1}{2}
+\frac{1}{4\sqrt{3}}\max
\left\{-\vp+7\vm-3\corr
+\frac{12}{\vp-7\vm+3\corr},
-2\vp+6\vm-2\corr+
\frac{6}{\vp-3\vm+\corr}
\right\},
\\
&u_{2}^{\max}:=
-1-\frac{4}{\sqrt{3}}(\vm-\corr),
\\
&u_{3}^{\min}:=-1+\frac{4}{\sqrt{3}}(\vm+r)
,
\\
&u_{3}^{\max}:=
\frac{1}{\sqrt{3}}\min
\left\{\vp -7\vm -3\corr
,	2\vp -6\vm -2\corr
\right\},
\end{split}
\end{align}
where $u_i^{\min}$ and $u_i^{\max}$, for $i\in\{1,2,3\}$, are the values of $u$ that correspond to
the boundary angles $\theta_i^{\min}$ and $\theta_i^{\max}$, respectively.
Note that the classical boundaries $(u_1^{\rm min}=u_3^{\rm max}=0$, $u_1^{\rm max}=u_2^{\rm min}=\infty$,
and $u_2^{\rm max}=u_3^{\rm min}=-1$) are automatically recovered in the limit
of vanishing relative moments.
However, it is important to point out that it is possible that, depending on the values of the moments,
$u_1^{\rm max}$ can be negative, and thus $u_1^{\rm max}<0<u_1^{\rm min}$.
This comes from the fact that the upper limit of the sector $S_1$, $\theta_1^{\rm max}$,
can be larger than its corresponding classical counterpart $4\pi/3$ (which corresponds to $u=\pm\infty$),
and in that case, according to Eq. \eqref{utheta}, it will be mapped to a large (in absolute value) negative number.
The sector $S_1$ will then be given not only by all the positive values of $u>u_1^{\rm min}$, but also by
the interval $(-\infty,u_1^{\rm max})$. Something similar happens with the lower limit of
the sector $S_2$: if $u_2^{\rm max}<0<u_2^{\rm min}$, the sector
$S_2$ will be defined by the two disjoint intervals $(-\infty,u_2^{\max})\cup(u_{2}^{\min},+\infty)$.

In summary, taking all these considerations into account, the ranges of values of the parameter $u$ that
define each sector $S_i$, for $i\in\{1,2,3\}$, will be denoted as $U_{S_i}$ and
are explicitly given as follows: 
\begin{align}
\label{quant_sectors_u}
\begin{split}
U_{S_1}&=
\begin{cases}
\left(u_{1}^{\min},u_1^{\max}\right),\quad\text{if}\quad u_1^{\max}>0,
\\
(-\infty,u_1^{\max})\cup(u_{1}^{\min},+\infty),\quad\text{if}\quad u_1^{\max}<0,
\end{cases}
\\
U_{S_2}&=
\begin{cases}
\left(u_{2}^{\min},u_2^{\max}\right),\quad\text{if}\quad u_2^{\min}<0,
\\
(-\infty,u_2^{\max})\cup(u_{2}^{\min},+\infty),\quad\text{if}\quad u_2^{\min}>0,
\end{cases}
\\
U_{S_3}&=\left(u_{3}^{\min},u_{3}^{\max}\right).
\end{split}
\end{align}
These sets $U_{S_i}$ will be used in Sec. \ref{subsec:quantum_u_map} to give the quantum transition map for the parameter $u$.

\subsection{The quantum Kasner map}\label{subsec:quantum_kasner_map}
Once the quantum sectors have been defined and the Kasner regimes are fully characterized by the
14 parameters $c_{\pm}$, $p_{\pm}$, and $K_{ijkl}$ [Eq. \eqref{constans_K_def}],
our goal is to obtain the transition rule that will provide the parameters of the postbounce
Kasner epoch in terms of those of the prebounce regime.
For this purpose, we will follow the same procedure that we have applied to the classical system in Sec. \ref{subsec:bounce}; i.e., we will compute the conserved quantities of the system,
then we will obtain the equations that relate the prebounce with the postbounce parameters
just by requesting that these quantities indeed have the same value before and after the transition,
and we will finally solve these equations. In addition, we will first consider a bounce
that takes place in sector $S_1$, by approximating the potential
$U$ by the Bianchi II potential $V=e^{-8\beta_+}/6$, and then obtain the transition law
for a bounce in the other two sectors $S_2$ and $S_3$ by considering appropriate rotations.

In the full quantum scenario, since the dynamics is described by an
infinite system of equations, there is an infinite amount of conserved quantities. However, if
four independent conserved operators $\hat{C}_1$, $\hat{C}_2$, $\hat{C}_3$, and $\hat{C}_4$ were known,
one would be able to generate an infinite set of conserved quantities just by taking the expectation
values of any product between them:
\begin{align}
\label{quantum_const_motion}
\langle \hat{C}_i^m\hat{C}_j^n\hat{C}_k^r\hat{C}_l^s \rangle&=const., \quad
\text{for}
\quad m,n,r,s\in\mathbb{N}
\quad
\text{and}\quad i,j,k,l\in\{1,2,3,4\}.
\end{align}
Rewriting these expressions in terms of moments would provide, at each order,
exactly the same amount of variables as conserved quantities, and thus one would have
the implicit solution of the full dynamics at hand. The main obstacle is obviously
the construction of such conserved operators, which is a nontrivial process and highly
depends on the ordering of the basic operators on the Hamiltonian.
However, we are considering a semiclassical truncation
at second order in moments, which implies that all terms of the order $\hbar^{3/2}$, and higher-order
powers, are negligible. Therefore, up to this level of approximation, the mentioned conserved
operators can be constructed just by promoting the conserved classical quantities  to operators, which in the case of a bounce in sector $S_1$ are given by Eq.
\eqref{conserved_quantities}. Thus, for a bounce in this specific sector (for the other two, $S_2$ and $S_3$, one just needs to implement the appropriate rotations), the conserved operators are given as follows:
\begin{eqnarray}\label{conserved_quantities_operatoC_3_1}
\hat{C}_1&:=&\hat{p}_-,
\\
\label{conserved_quantities_operatoC_3_2}
\hat{C}_2&:=&2\hat H-\hat p_+,
\\
\label{conserved_quantities_operatoC_3_3}
\hat{C}_3&:=&\left(\frac{2\hat{H}-\hat p_+}{\hat p_-}\right)\hat \beta_-+\hat\beta_+-2\alpha,
\end{eqnarray}
\begin{eqnarray}
\label{conserved_quantities_operatoC_3_4}
\hat{C}_4&:=&e^{\frac{4[(2\hat H-\hat p_+)^2-3\hat p_-^2]^{1/2}\hat \beta_-}{\hat p_-}}\left(
\frac{2}{1+\frac{2\hat p_+-\hat H}{[(2\hat H-\hat p_+)^2-3\hat p_-^2]^{1/2}}}-1
\right),
\end{eqnarray}
which will be assumed to be Weyl-ordered.
Note, however, that there might be explicit $\hbar$ terms in the conserved operators that
we are not considering here.
Nonetheless, when expanding the expectation values of the conserved operators and writing them in terms of the moments, such terms would multiply a moment, which would lead to a contribution of at least fourth order.
Therefore, even if we are just considering the second-order approximation, our analysis is very
general in the sense that it is valid for any ordering of the basic operators on the Hamiltonian.

As already commented above, at second order, the quantum Kasner regimes are characterized by 14 parameters. Hence, in order to completely solve
the quantum transition law, one needs to construct the same number of constants of motion. These would be
given by the four expectation values of the conserved operators,
\begin{eqnarray}
&&\langle \hat{C}_i \rangle,\hspace{0.5cm}i\in\{1,2,3,4\},
\label{expr1}
\end{eqnarray}
their four fluctuations,
\begin{eqnarray}
&&\langle (\hat{C}_i-C_i)^2 \rangle,\hspace{0.5cm}i\in\{1,2,3,4\},
\label{expri_2}
\end{eqnarray}
and the six crossed correlations,
\begin{eqnarray}
&&\langle (\hat{C}_1-C_1) (\hat{C}_2-C_2)+(\hat{C}_2-C_2)(\hat{C}_1-C_1) \rangle,
\label{expr12}
\\
&&\langle (\hat{C}_1-C_1) (\hat{C}_3-C_3)+(\hat{C}_3-C_3)(\hat{C}_1-C_1) \rangle,
\label{expr13}
\\
&&\langle (\hat{C}_1-C_1) (\hat{C}_4-C_4)+(\hat{C}_4-C_4)(\hat{C}_1-C_1) \rangle,
\label{expr14}
\\
&&\langle (\hat{C}_2-C_2) (\hat{C}_3-C_3)+(\hat{C}_3-C_3)(\hat{C}_2-C_2) \rangle,
\label{expr23}
\\
&&\langle (\hat{C}_2-C_2) (\hat{C}_4-C_4)+(\hat{C}_4-C_4)(\hat{C}_2-C_2) \rangle,
\label{expr24}
\\
&&\langle (\hat{C}_3-C_3) (\hat{C}_4-C_4)+(\hat{C}_4-C_4)(\hat{C}_3-C_3) \rangle,
\label{expr34}
\end{eqnarray}
where we have defined $C_i:=\langle\hat C_i\rangle$ for $ i\in\{1,2,3,4\}$. The next step is to
write these expressions in terms of the moments $\Delta(\beta_+^ip_+^j\beta_-^kp_-^l)$
by performing an expansion around the expectation values $\beta_{\pm}$ and $p_{\pm}$,
and then truncate the series at second order.
Note, in particular, that we have defined the above expressions as expectation
values of powers of differences between the operator and its corresponding expectation value, like $(\hat{C}_i-C_i)^2$, instead of expectation values
of their powers, like $\hat{C}_i^2$. By doing so, together with the chosen completely symmetric ordering of the operators $\hat{C}_i$, one automatically gets pure second-order real expressions, since there are no zeroth-order contributions when expanding such expectation values.

Once we have obtained these 14 constants of motion in terms of the expectation values and second-order
moments, we evaluate them in each Kasner epoch by considering the behavior of each variable:
in this regime, the momenta
$p_{\pm}$ and their fluctuations $\Delta(p_+^ip_-^j)$ are constant, whereas
the evolution of the rest of the variables is given in Eqs. \eqref{quantum_kasner_beta_+_full}--\eqref{quantum_kasner_beta_-_full} and Eqs. \eqref{quantum_kasner_delta_beta_+_p_+}--\eqref{quantum_kasner_delta_beta_+_beta_-}. Then, the requirement of the conservation of these quantities leads to a system of equations
that relates the parameters of the prebounce Kasner epoch $(\overline c_{\pm}, \overline p_{\pm}, \overline{K}_{ijkl})$
with those of the postbounce one $(\widetilde c_{\pm}, \widetilde p_{\pm}, \widetilde{K}_{ijkl})$. This system of equations is very involved and the complete solutions are given
in Appendix \ref{app:explicit}. These solutions are the generalization of the classical function $B$ [Eq. \eqref{bounce}]
for a bounce in the sector $S_1$,
and they represent one of the main results of this paper. In the following, we will
comment on the specific transition laws for certain variables and their general structure, and then obtain
the complete quantum Kasner map $T_Q$ for bounces in any sector.

Let us first analyze change of parameters that also appear in the classical
system---that is, the generalization
of the classical map [Eqs. \eqref{transition_p_-}--\eqref{transition_c_-}].
After a bounce in sector $S_1$, these parameters are modified as follows:
\begin{align}
\label{trans_general_p_-}
\widetilde{p}_-&=\overline{p}_-,
\\
\label{trans_general_p_+}
\widetilde{p}_+&=
\frac{1}{3}\left(4\overline P-5\overline p_+\right)
+\frac{2}{3\overline P^3}
\left(
\overline p_+^2 \overline{\Delta(p_-^2)}-2p_-p_+\overline{\Delta(p_+p_-)}
+p_-^2\overline{\Delta(p_+^2)}
\right),
\end{align}
\begin{align}
\nonumber
\widetilde{c}_+&=
B_+(\overline{c}_+,\overline{p}_+,\overline{p}_-)+
\frac{1}{2}\left(
\dfrac{\partial^2B_+}{\partial\overline p_+^2}\overline{\Delta(p_+^2)}+
\dfrac{\partial^2B_+}{\partial\overline p_-^2}\overline{\Delta(p_-^2)}
\right)+\dfrac{\partial^2B_+}{\partial\overline c_+\partial\overline p_+}\overline{k}_{\scriptsize_{++}}
\\
&
+\dfrac{\partial^2B_+}{\partial\overline c_+\partial\overline p_-}\overline{k}_{\scriptsize_{+-}}+
\dfrac{\partial^2B_+}{\partial\overline p_+\partial\overline p_-}\overline{\Delta(p_+p_-)},
\label{trans_general_c_+}
\\
\nonumber
\widetilde{c}_-&=B_-(\overline{c}_+,\overline{p}_+,\overline{c}_-,\overline{p}_-)+
\frac{1}{2}\left(
\dfrac{\partial^2B_-}{\partial\overline p_+^2}\overline{\Delta(p_+^2)}+
\dfrac{\partial^2B_-}{\partial\overline p_-^2}\overline{\Delta(p_-^2)}
\right)+\dfrac{\partial^2B_-}{\partial\overline c_+\partial\overline p_+}\overline{k}_{\scriptsize_{++}}
\\
&
+\dfrac{\partial^2B_-}{\partial\overline c_+\partial\overline p_-}\overline{k}_{\scriptsize_{+-}}+
\dfrac{\partial^2B_-}{\partial\overline p_+\partial\overline p_-}\overline{\Delta(p_+p_-)},
\label{trans_general_c_-}
\end{align}
where $B_+$ and $B_-$ are the classical transition laws for $c_+$ and $c_-$
defined in Eqs. \eqref{transition_c_+} and \eqref{transition_c_-}.
First of all, we observe from Eq. \eqref{trans_general_p_-} that the transition law for $p_-$
remains unchanged from the classical one, as $p_-$ is itself a constant of motion.
On the contrary, quantum corrections appear in the transition law for the rest of the variables.
However, only pure fluctuations of the momenta $\Delta(p_{\pm}^2)$ contribute
to the transition law for $p_+$ [Eq. \eqref{trans_general_p_+}], whereas
from Eqs. \eqref{trans_general_c_+} and \eqref{trans_general_c_-},
one can see that the final values of $c_{\pm}$, which characterize the evolution of the
shape parameters,
get quantum corrections that depend only on initial parameters unrelated to $\beta_-$,
that is, $\Delta(p_+^ip_-^j)$ or $k_{\scriptsize_{\pm\pm}}$.
Note also that in particular, these last two transition laws are written as a linear combination
of derivatives of the classical ones, $B_+$ and $B_-$. This is a general feature at
this order of truncation, and one can indeed obtain the quantum transition rules
as formal expansions of the corresponding classical ones. This
alternative method is quite straightforward in order to obtain the
quantum version of the classical relations in Eqs. \eqref{transition_p_-}--\eqref{transition_c_-}.
However, the calculation of transition laws for fluctuations and correlations is far from trivial, as it
requires the summation and composition of several series. Therefore, even if we have also performed
the computation with this alternative method and obtain similar
results, we will not provide more details here, since it would not
contribute to enlightening the discussion. 

Then, regarding the pure fluctuations and correlation of the momenta $p_{\pm}$,
their transition laws after a bounce in sector $S_1$ are given as follows:
\begin{align}
\label{trans_general_delta_p_-_2}
\widetilde{\Delta(p_-^2)}&=\overline{\Delta(p_-^2)},
\\
\label{trans_general_delta_p_+_p_-}
\widetilde{\Delta(p_+p_-)}&=
\frac{1}{3\overline P}
\left[
4\overline p_-\overline{\Delta(p_-^2)}-(5\overline P-4\overline p_+)\overline{\Delta(p_+p_-)}
\right],
\\
\label{trans_general_delta_p_+_2}
\widetilde{\Delta(p_+^2)}&=
\frac{1}{9\overline P^2}
\left[16\overline p_-^2
\overline{\Delta(p_-^2)}-8\overline p_-(5\overline P-4\overline p_+)
\overline{\Delta(p_+p_-)}+(5\overline P-4\overline p_+)^2
\overline{\Delta(p_+^2)}
\right].
\end{align}
As expected, the fluctuation of $p_-$ is conserved, and its value enters the
transition laws for the other two moments. In particular, the more coupled
of these relations is the transition law for the fluctuation of $p_+$, whose final value
depends on its own initial value as well as those of $\Delta(p_-^2)$ and $\Delta(p_+p_-)$.

The transition laws for the rest of the parameters are quite long and complicated,
so we refer the reader to Appendix \ref{app:explicit} for the explicit expressions.
However, as a general feature, they can be schematically written in the following way:
\begin{align}\label{trans_gen_moments}
\widetilde{K}_{ijkl}=
\sum_{m=0}^{i+k}\sum_{n=0}^{i+j+k}
\sum_{r=0}^k\sum_{s=l}^{2}
a_{mnrs}\,
\overline{K}_{mnrs},
\end{align}
with certain coefficients $a_{mnrs}$, which depend on the expectation values,
and the coefficients $K_{ijkl}$ defined in Eq. \eqref{constans_K_def}.
That is, the final parameter $\widetilde{K}_{ijkl}$ only depends on the values of the initial parameters $\overline{K}_{mnrs}$, with $m\leq i+k$, $n\leq i+j+k$, $r\leq k$, and $s\geq l$. 

In summary, these transition laws define the quantum bounce map $B_Q$, which is the quantum
version of the classical bounce $B$ defined in Eq. \eqref{bounce}. This map provides directly the quantum transition law when the system bounces in sector $S_1$. However, when the system bounces
against the walls located in sectors $S_2$ or $S_3$, a rotation is also required, as explained in Sec. \ref{subsec:bounce}.
Since the angle of the velocity vector $\theta$ defines in which sector the bounce will occur, as shown in Eq. \eqref{quant_sectors},
the quantum Kasner map is explicitly written as follows:
\begin{align}\label{trans_quantum}
T_Q\left(c_+,c_-,p_+,p_-, \{{K}_{ijkl}\}\right)=\begin{cases}
B_Q\left(c_+,c_-,p_+,p_-, \{{K}_{ijkl}\}\right), &\theta\in(\theta_1^{\min},\theta_1^{\max}),
\\
R_Q^{-1}\circ B_Q\circ R_Q\left(c_+,c_-,p_+,p_-, \{{K}_{ijkl}\}\right), &\theta\in(\theta_2^{\min},\theta_2^{\max}),
\\
R_Q\circ B_Q\circ R_Q^{-1}\left(c_+,c_-,p_+,p_-, \{{K}_{ijkl}\}\right), &\theta\in(\theta_3^{\min},\theta_3^{\max}),
\end{cases}
\end{align}
where $R_Q$ represents the action produced by a $2\pi/3$ clockwise rotation
in the plane of the shape parameters on the different variables.
Specifically, on the classical parameters
$(c_+, c_-,p_+,p_-)$, the linear function $R_Q$ acts as $R$ [Eq. \eqref{C_4_dim}], whereas its action on the parameters ${K}_{ijkl}$ [Eq. \eqref{constans_K_def}] is given by
\begin{align}\nonumber
{K}_{ijkl}\stackrel{R_Q }{\longmapsto}
&
\sum_{m=0}^i\sum_{n=0}^j\sum_{r=0}^k\sum_{s=0}^l
\binom{i}{m}\binom{j}{n}\binom{k}{r}\binom{l}{s}
\frac{\left(-1\right)^{k+l+m+n}}{2^{i+j+k+l}}
\left(\sqrt{3}\right)^{i+j+r+s-m-n}
K_{m+r,n+s,i+k-m-r,j+l-n-s}.
\end{align}
Likewise, $R_Q^{-1}$ stands for the action of the $2\pi/3$ counterclockwise rotation,
and its explicit form is the same as that of $R_Q$ above, just with the replacement $\sqrt{3}\rightarrow -\sqrt{3}$.
Finally, let us recall that for values of $\theta$ in the exit channels
		$(\theta_1^{\rm max},\theta_2^{\rm min})\cup(\theta_2^{\rm max},\theta_3^{\rm min})\cup (\theta_3^{\rm max},\theta_1^{\rm min})$
		the system will
		not suffer any transition
		and the corresponding Kasner map is the identity.

\subsection{The quantum $u$-map}
\label{subsec:quantum_u_map}

From the previous results, we can now derive the quantum version of the transition law for the parameter $u$
 \eqref{trans_u_general}. For this purpose, following the same procedure as in Sec. \ref{subsec:bounce},
we just need to construct the transition law that relates the prebounce and postbounce angles $\theta$
and then use the relation in Eq. \eqref{utheta}. Since the angle $\theta$
is determined by the ratio $p_-/p_+$, it is clear from relations \eqref{trans_general_p_-} and \eqref{trans_general_p_+} that its transition
law will be coupled only to the transition laws for the fluctuations $\Delta(p_+^2)$ [Eq. \eqref{trans_general_delta_p_+_2}] and $\Delta(p_-^2)$
[Eq. \eqref{trans_general_delta_p_-_2}],
and for the correlation $\Delta(p_+p_-)$ [Eq. \eqref{trans_general_delta_p_+_p_-}].
Making use of the same parametrization introduced above for these moments---i.e.,
$\vp:=\sqrt{3}\Delta(p_+^2)/(8P^2)$, $\vm:=\Delta(p_-^2)/(8\sqrt{3}P^2)$,
and $r:=\Delta(p_+p_-)/(4P^2)$---one obtains the following coupled system that maps
the prebounce values $(\overline u,\overline v_+,\overline v_-,\overline r)$ to the postbounce values
$(\widetilde u,\widetilde v_+,\widetilde v_-,\widetilde r)$:
\begin{align}
\nonumber
\bullet\;&\text{If } \overline u\in U_{S_1}:
\\
\widetilde{u}&=
-\overline u-
\frac{2(\overline u^2-1)}{\sqrt{3}(1+\overline u^2+\overline u^4)}
\left[
(\overline u^2-1)^2\overline v_+
+(1+4\overline u+\overline u^2)^2\overline v_-
-(\overline u^2-1)(1+4\overline u+\overline u^2)\overline r
\right],
\label{transition_u_quantum_S1}
\\
\widetilde{v}_+&=\overline v_++
\frac{2(\overline u^2-1)}{1-\overline u+\overline u^2}\left[\frac{2(\overline u^2-1)}{1-\overline u+\overline u^2}\overline v_-
-\overline r
\right],
\label{trans_vp_S1}
\\
\widetilde{v}_-&=
\frac{(1+\overline u+\overline u^2)^2}{(1-\overline u+\overline u^2)^2}\overline v_-,
\label{trans_vm_S1}
\\
\widetilde{r}&=
\frac{1+\overline u+\overline u^2}{1-\overline u+\overline u^2}\left[\frac{4(\overline u^2-1)}{(1-\overline u+\overline u^2)}\overline v_-
-\overline r
\right].
\label{trans_r_S1}
\\
\bullet\;&\text{If } \overline u\in U_{S_2}:
\nonumber
\\
\widetilde{u}&=
-\overline{u}-2+\frac{2\overline u(\overline u+2)
	\left[
	(\overline u^2-1)^2\overline v_+
	+(1+4\overline u+\overline u^2)^2\overline v_-
	-(\overline u^2-1)(1+4\overline u+\overline u^2)\overline r
	\right]
}{\sqrt{3}(1+\overline u+\overline u^2)
	(3+3\overline u+\overline u^2)},
\label{transition_u_quantum_S2}
\\
\widetilde{v}_+&=\frac{1}{(3+3\overline u+\overline u^2)^2}\left[
\overline u^2\overline v_++
(3+2\overline u+\overline u^2)^2\overline v_-
-\overline u(3+2\overline u+\overline u^2)\overline r
\right],
\label{trans_vp_S2}
\\
\widetilde{v}_-&=\frac{1}{(3+3\overline u+\overline u^2)^2}\left[
(1+\overline u)^4\overline v_++
(2+\overline u)^2\overline v_-
-(1+\overline u)^2(2+\overline u)\overline r
\right],
\label{trans_vm_S2}
\\
\widetilde{r}&=-\frac{1}{(3+3\overline u+\overline u^2)^2}\left[
2\overline u(1+\overline u)^2\overline v_++
2(2+\overline u)(3+2\overline u+\overline u^2)\overline v_-
-(\overline u^4+4 \overline u^3+9 \overline u^2+10 \overline u+3)\overline r
\right].
\label{trans_r_S2}
\\
\bullet\;&\text{If } \overline u\in U_{S_3}:
\nonumber
\\
\widetilde{u}&=-\frac{\overline u}{1+2\overline u}
-\frac{2\left[
	(\overline u^2-1)^2\overline v_+
	+(1+4\overline u+\overline u^2)^2\overline v_-
	-(\overline u^2-1)(1+4\overline u+\overline u^2)\overline r
	\right]}{\sqrt{3}(1+2\overline u)(1+\overline u+\overline u^2)(1+3\overline u+3\overline u^2)}
,
\label{transition_u_quantum_S3}
\\
\widetilde{v}_+&=
\frac{1}{	(1+3\overline u+3\overline u^2)^2}\left[
\overline u^2\overline v_++
(1+2\overline u+3\overline u^2)^2\overline v_-
+\overline u(1+2\overline u+3\overline u^2)\overline r
\right]
,
\label{trans_vp_S3}
\\
\widetilde{v}_-&=
\frac{1}{	(1+3\overline u+3\overline u^2)^2}\left[
(1+\overline u)^4\overline v_+ +\overline u^2(1+2\overline u)^2\overline v_-
+u(1+\overline u)^2(1+2\overline u)\overline r
\right]
,
\label{trans_vm_S3}
\\
\widetilde{r}&=
\frac{1}{(1+3\overline u+3\overline u^2)^2}\left[2\overline u
(1+\overline u)^2\overline v_+ +2\overline u(1+4\overline u+7\overline u^2+6\overline u^3)\overline v_-
+(1+4\overline u+9\overline u^2+10\overline u^3+3\overline u^4)\overline r
\right]
.
\label{trans_r_S3}
\end{align}
The sets $U_{S_1}$, $U_{S_2}$, and $U_{S_3}$, which characterize the different sectors,
have been defined in Eq. \eqref{quant_sectors_u},
and they depend on all the prebounce values of the relative quantum moments $(\overline{v}_+, \overline{v}_-, \overline{r})$.
For values of $\overline u$ in the exit channels
$ {U}_{S_1}^{c}\cap{U}_{S_2}^{c}\cap{U}^{c}_{S_3}$ (where the superindex $c$ denotes the complement set)
the transition law is the identity map. Note that, despite the complicated forms of some of the denominators,
none of them have real roots, except for the factor $(1+2\overline u)$ that appears in Eq. \eqref{transition_u_quantum_S3}.
This divergence at $\overline u=-1/2$ is expected, since it already appears in the corresponding classical relation
[Eq. \eqref{trans_u_general}]
and simply maps the value $\overline u=-1/2$ to $\widetilde u=\infty$.

Let us briefly comment on the transition law for the parameter $u$
given by Eqs. \eqref{transition_u_quantum_S1}, \eqref{transition_u_quantum_S2}, and \eqref{transition_u_quantum_S3}.
One can observe in these relations that the quantum effects are completely encoded in
the linear contributions of the moments
$v_+$, $v_-$, and $r$. These terms have different signs and, depending on the specific values of the moments,
the quantum effects might either increase or decrease the final value $\widetilde u$.
On the one hand, even if the fluctuations $\overline v_{\pm}$ are positive definite, their coefficients
in these relations are not, and thus the sign of their contribution depends on the prebounce value of $u$.
More precisely, they are positive for the set of points $\{\overline u\in U_{S_1}\mid \overline u>1\}$, $\{\overline u\in U_{S_2}\mid \overline u<-2\}$, and  $\{\overline u\in U_{S_3} \mid \overline u<-1/2\}$,
while they are negative for $\{\overline u\in U_{S_1}\mid \overline u<1\}$, $\{\overline u\in U_{S_2}\mid \overline u>-2\}$, and  $\{\overline u\in U_{S_3} \mid \overline u>-1/2\}$. On the other hand, the correlation $\overline r$ does not have a predefined sign, and moreover, the coefficients that multiply it can be either positive or negative.
For instance, for the $S_1$ sector ($\overline u\in U_{S_1}$) its coefficient is positive definite, and thus this
quantum correction always contributes with the same sign as the correlation itself.
Nevertheless, for the other two sectors, $S_2$ ($\overline u\in U_{S_2}$) and $S_3$ ($\overline u\in U_{S_3}$),
its coefficients are positive only for the intervals $\overline u\in(-2-\sqrt{3},-2)$ and $\overline u\in(-1/2,-2+\sqrt{3})$, respectively; consequently, these contributions will depend on the sign of $\overline{r}$ as well as on the prebounce value $\overline u$. Finally, it is interesting to note that for the prebounce values
$\overline u\in\{-2, -1/2, 1\}$, there are no quantum corrections in the transition law for the parameter $u$,
regardless of the value of the relative moments. In the classical model, these values correspond to the last
bounce of the system, as they are mapped to the fixed (exit) points $\{0,\infty,-1\}$ of the map [Eq. \eqref{trans_u_general}].
However, in this quantum case, their image might no longer be an exit point.

Once the classical $u$-map [Eq. \eqref{trans_u_general}] was derived, in Sec. \ref{subsec:sequence_bounces} we studied a sequence of bounces against the potential walls and obtained certain explicit properties. However, an equivalent study for the quantum case is much more involved, since
one has to deal with the coupled multiparameter map [Eqs. \eqref{transition_u_quantum_S1}--\eqref{trans_r_S3}] which, even if it is linear
in the relative moments $(\vp,\vm,r)$, is highly nonlinear in the parameter $u$.
This map encodes the stochastic properties of the quantum model, and it explicitly shows that the evolution
of a general point after a sequence of bounces will probably be very different from its classical counterpart,
as quantum effects might pile up during different bounces and produce a highly deformed trajectory.
As explained in Sec. \ref{subsec:quantum_sectors}, some conclusions regarding the stochastic properties of the quantum model can already be inferred without a more detailed analysis:
since quantum effects widen the exit channels---they are no longer the fine-tuned values $u=0,-1$, or $\infty$, as shown in \eqref{quant_sectors_u}---given an initial set of points that lies in one of these channels, they will all follow qualitatively the same dynamics, and thus, at this scale, the chaotic behavior will cease.
Still, the open question is whether all points end up in one of these exit channels,
which would imply that for generic initial data, the bouncing behavior of the quantum Mixmaster model
will eventually stop, and the system will enter a final Kasner regime until the singularity.
In this case, the chaotic repeller of the classical model would be removed by the quantum effects
and would not exist for the quantum model.
However, in order to check this issue, a detailed iterative analysis of the map would have to be performed,
which is beyond the scope of the present paper.

\section{Analysis of specific quantum states}
\label{sec:specific_state}
In the previous section, we have obtained the quantum Kasner map [Eq. \eqref{trans_quantum}] for a generic state.
Since the transition rules are very complicated and highly sensitive on the prebounce
state, it is not possible to draw generic qualitative conclusions about the impact
of the bounce on the state. Therefore, in order to gain some intuition, in this section
we will consider certain specific prebounce quantum states and analyze how the bounce changes
their properties. More precisely, in Sec. \ref{subsec:uncorrelated} we will assume an initially (at $\alpha=0$)
uncorrelated state and study certain features of the postbounce state by using the analytic
quantum Kasner map [Eq. \eqref{trans_quantum}] obtained above. Furthermore, in Sec. \ref{subsec:dynamical_evolution} we will consider the numerical resolution
of the full equations of motion [Eqs. \eqref{quantum_em_beta_+}--\eqref{quantum_em_moments}] for an initial Gaussian state. This analysis will serve as
a test of the main assumptions used to derive the quantum Kasner map
(particularly that the quantum system asymptotically follows a sequence of Kasner epochs
joined by quick transitions), as well as a test of the map itself.

\subsection{The quantum Kasner map for an initially uncorrelated state}\label{subsec:uncorrelated}

Let us assume a quantum state that at $\alpha=0$ is following a Kasner epoch and all its correlations
are vanishing. This implies that the characteristic quantum parameters of this Kasner epoch are
all zero, except for $\Delta(p_+^2)$, $\Delta(p_-^2)$, ${c}_{\scriptsize_{++}}$, and
${c}_{\scriptsize_{--}}$.
Note, however, that such a state is not a coherent state, and the Kasner dynamics itself will generate
nonvanishing correlations as time goes by, as given by the relations \eqref{quantum_kasner_delta_beta_+_p_+}--\eqref{quantum_kasner_delta_beta_+_beta_-}. In this case, the quantum Kasner map simplifies a bit, and it is possible to obtain certain bounds
and specific properties for the moments of the postbounce state. For the rest of the section, let us
focus specifically on bounces that happen in $S_1$.

\begin{figure}[t]
	\centering
	\includegraphics[width=0.6\linewidth]{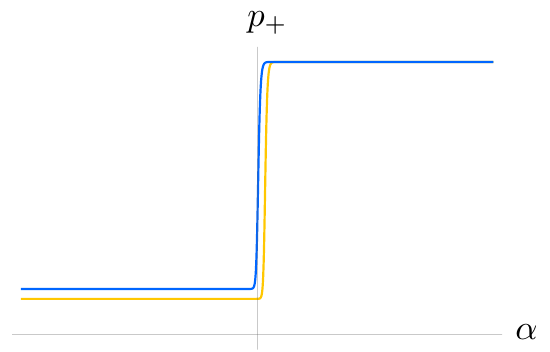}
	\caption{Comparison of the classical and quantum evolution of the momentum $p_+$ during a bounce in sector $S_1$. The yellow line represents the classical evolution and the blue line represents the quantum
		evolution.}
	\label{fig:gaussian_p}
\end{figure}

In particular, apart from the properties obeyed by any generic quantum state,
such as the momentum $p_-$ being conserved through the bounce,
one can see from the transition law for $p_+$ [Eq. \eqref{trans_general_p_+}], which in this case takes the form
\begin{equation}
 \widetilde{p}_+=\frac{1}{3}\left(4\overline P-5\overline p_+\right)
+\frac{2}{3\overline P^3}
\left(
\overline p_+^2 \overline{\Delta(p_-^2)}
+p_-^2\overline{\Delta(p_+^2)}
\right),
\end{equation}
that all the quantum contributions to the postbounce value of the momentum $p_+$ are positive.
Therefore, for such uncorrelated states, the postbounce value of $p_+$ is larger than in
the classical system. This feature can be observed in Fig. \ref{fig:gaussian_p}, where
the classical and quantum evolutions of $p_+$ are shown.

\begin{figure}[t]
	\centering
	\begin{subfigure}[h]{0.5\textwidth}
		\centering
		\includegraphics[width=\textwidth]{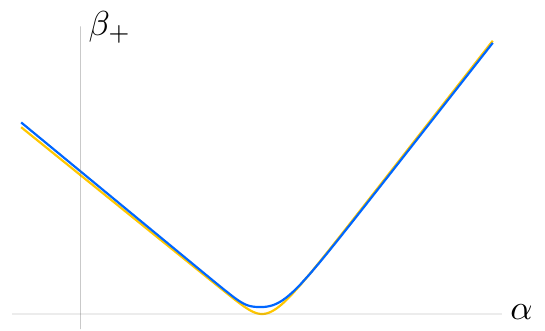}
	\end{subfigure}
	\hfill
	\begin{subfigure}[h]{0.49\textwidth}
		\centering
		\includegraphics[width=\textwidth]{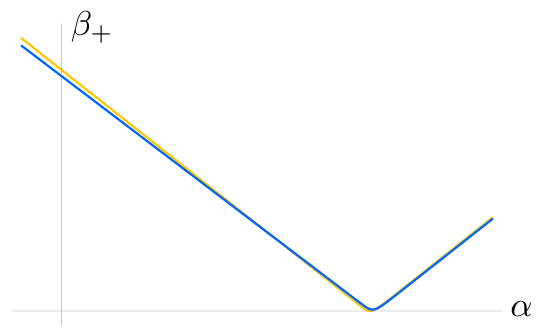}
	\end{subfigure}
	\caption{Comparison of the classical and quantum evolution of
				$\beta_{+}$ during a bounce in sector $S_1$. The yellow line represents the classical evolution and the blue one the quantum
				evolution. Each of the plots corresponds to different initial conditions. In the first plot, the quantum effects
				decrease the value of the $\widetilde{c}_+$ with respect to its classical counterpart, whereas in the second plot,
				they produce an enhancement.}
	\label{fig:gaussian_beta_+}
\end{figure}

Concerning the parameters $c_{+}$ and $c_{-}$, their transition laws are much more involved,
and it is not possible to obtain such generic results. In fact, quantum moments can either increase
or decrease the postbounce values of these parameters as compared to their classical counterparts,
depending on the specific initial (prebounce) conditions.
This can be seen in Fig.  \ref{fig:gaussian_beta_+}, where the transition
of the shape parameters is plotted for different initial
data, and the parameter $\widetilde{c}_{+}$ can be read from the corresponding vertical intercept. 

Regarding the fluctuations of the momenta $p_{\pm}$, as for any quantum state,
the fluctuation $\Delta(p_-^2)$ remains unchanged during any transition on the sector $S_1$,
while for this uncorrelated state, the change of the fluctuation $\Delta(p_+^2)$ is given by
\begin{align}
&\widetilde{\Delta(p_+^2)}=\left(\frac{5\overline P-4\overline p_+}{3\overline P}\right)^2\overline{\Delta(p_+^2)}+
\left(\frac{4\overline p_-}{3\overline P}\right)^2
\overline{\Delta(p_-^2)}.
\label{gaus_transition_delta_p_+_2}
\end{align}
From this expression, and since for bounces in sector $S_1$ the prebounce momentum $\overline{p}_+$ is bounded by
$\overline{P}/2< \overline{p}_+\leq\overline{P}$,
one can deduce that the postbounce value of this fluctuation is in the following range:
\begin{align}\label{fluct_p_+_gauss}
\frac{1}{9} \overline{\Delta(p_+^2)}\leq\;
\widetilde{\Delta(p_+^2)}<
\overline{\Delta(p_+^2)}
+	\frac{4}{3}\overline{\Delta(p_-^2)}.
\end{align}
Thus, we conclude that, after a bounce in sector $S_1$, the initial uncorrelated state can either be stretched or compressed in
the $p_+$ direction, depending on the initial values of $p_{+}$ and $p_-$.
On the contrary, one can see from Eqs. \eqref{quantum_transition_delta_beta_+_2} and \eqref{quantum_transition_delta_beta_-_2}, which are largely simplified in this case,
that the parameters that characterize the fluctuation of the shape parameters $c(\beta_{\pm}^2)$
$c_{++}$ and $c_{--}$ always get amplified by bounces in sector $S_1$---that is, 
\begin{align}\label{fluct_beta_gauss}
\widetilde{c}_{\scriptsize_{++}}>\overline{c}_{\scriptsize_{++}},
\qquad 
\widetilde{c}_{\scriptsize_{--}}>\overline{c}_{\scriptsize_{--}}.
\end{align}

Furthermore, from Eq. \eqref{quantum_transition_delta_p_+_p_-}, one can easily see that the postbounce correlation
$\widetilde {\Delta(p_+p_-)}$ has the same sign as the momentum $\overline p_-$
and, taking again into account the relation
$\overline{P}/2< \overline{p}_+\leq\overline{P}$ for bounces in sector $S_1$,
it is bounded by
\begin{align}\label{corC_p_gaus}
&
\lvert\widetilde{\Delta(p_+p_-)}\rvert<
\frac{2}{\sqrt{3}} \overline{\Delta(p_-^2)}.
\end{align}
That is, the larger the value of $\overline{\Delta(p_-^2)}$, the stronger the correlation
that can be built between $p_+$ and $p_-$ after the bounce.
However, the transition laws [Eqs. \eqref{quantum_transition_delta_beta_+_p_-}--\eqref{quantum_transition_delta_beta_-_p_+}]
and \eqref{quantum_transition_delta_beta_+_beta_-} for the rest of the initially vanishing parameters that describe the correlations are still very involved for this particular case.
All of them have a strong sensitivity on the initial values, and it is not possible to provide
bounds for the corresponding postbounce parameters. The main technical reason is that, in the mentioned laws,
there appears a term of the form $8 \overline c_++\ln\left[\frac{2}{3} (2 \overline p_+-\overline P)^2\right]$,
which is unbounded and can take any real value.
Still, we can state that at least one of the parameters that describe the correlations
will always be nonzero after a bounce in sector $S_1$. Indeed, the postbounce correlation
$\widetilde {\Delta(p_+p_-)}$ can only vanish as long as the initial momentum $\overline{p}_-$
is also vanishing and, in such a case,
\begin{align}
\label{correlations_p_-_0}
\widetilde{k}_{\scriptsize_{++}}=\frac{1}{3\overline p_+}\overline{\Delta(p_+^2)},
\end{align}
which is positive. Therefore, if the prebounce state is uncorrelated at $\alpha=0$,
the bounce will modify it, and generically the postbounce state will not be of this type.

\subsection{Dynamical evolution for an initial Gaussian state}\label{subsec:dynamical_evolution}

\begin{figure}[t]
	\centering
	\begin{subfigure}[h]{0.48\textwidth}
		\centering
		\includegraphics[width=\textwidth]{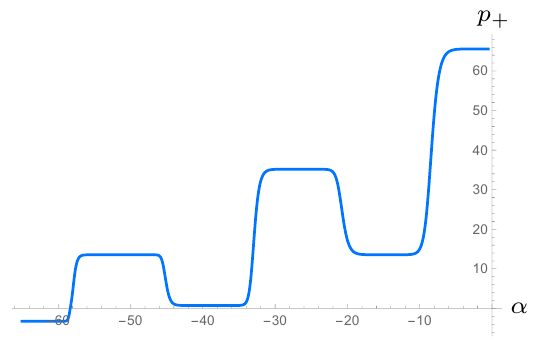}
	\end{subfigure}
	\hfill
	\begin{subfigure}[h]{0.48\textwidth}
		\centering
		\includegraphics[width=\textwidth]{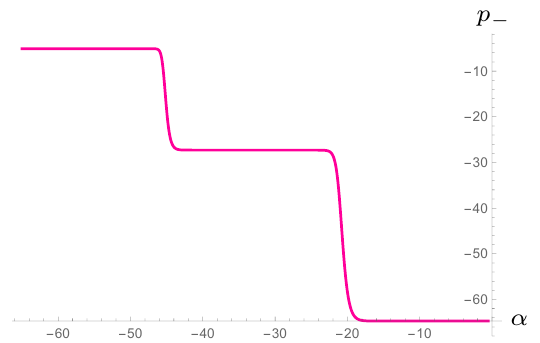}
	\end{subfigure}
	\caption{Dynamical evolution of the momenta $p_+$ and $p_-$ for an initial Gaussian state.}
	\label{fig:p_gaussian}
	\centering
	\begin{subfigure}[h]{0.48\textwidth}
		\centering
		\includegraphics[width=\textwidth]{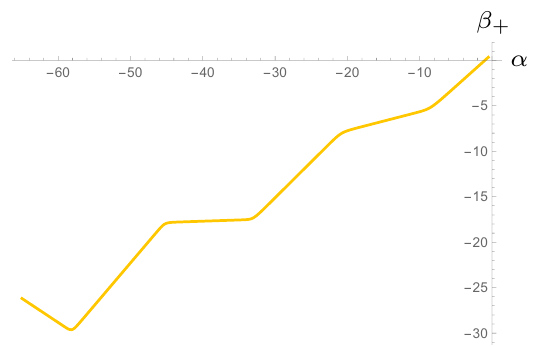}
	\end{subfigure}
	\hfill
	\begin{subfigure}[h]{0.48\textwidth}
		\centering
		\includegraphics[width=\textwidth]{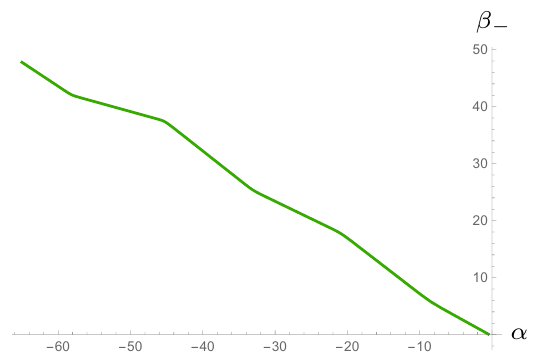}
	\end{subfigure}
	\caption{Dynamical evolution of the shape parameters $\beta_+$ and $\beta_-$ for an initial Gaussian state.}
	\label{fig:beta_gaussian}
\end{figure}

In this section, we will briefly discuss the dynamical evolution for an initial Gaussian state,
mainly to show that, as with the classical model, asymptotically the quantum system follows a sequence of Kasner regimes connected by quick transitions. We will choose initial conditions at $\alpha=0$, and thus this
Gaussian state will be of the form of the uncorrelated states described above, with $K_{ijkl}=0$ for any odd index, and
the fluctuations of the basic variables given by
\begin{equation}
\label{initial_gaussian}
\overline{c}_{\scriptsize_{++}}=
\frac{\hbar^{2}}{2\sigma_{+}^2}
,\quad
\overline{\Delta(p_{+}^2)}=\frac{\sigma_{+}^{2}}{2},
\quad
\overline{c}_{\scriptsize_{--}}=
\frac{\hbar^{2}}{2\sigma_{-}^2},
\quad
\overline{\Delta(p_{-}^2)}=\frac{\sigma_{-}^{2}}{2},
\end{equation}
where $\sigma_+$ and $\sigma_-$ are the corresponding Gaussian widths.
Furthermore, the initial conditions for the expectation values will be chosen
so that the system begins in a Kasner regime.

The numerical study shows the expected behavior as the system evolves toward lower values of $\alpha$:
a sequence of Kasner regimes connected by quick transitions. These transitions take place during a
very short period of time, as compared to the duration of the epochs themselves,
and the lower the value of $\alpha$, the longer each Kasner regime persists.
These features can be seen in Figs. \ref{fig:p_gaussian} and \ref{fig:beta_gaussian},
where the evolution of the basic variables $(\beta_{\pm}, p_{\pm})$ has been plotted. As described in Sec. \ref{subsec:Kasner_quantum},
during the Kasner epochs, the momenta $p_{\pm}$ remain constant, whereas the shape parameters
$\beta_{\pm}$ are linear in $\alpha$. Moreover, the slopes of the shape parameters observed
from these figures are in accordance with the expected ones in Eqs. \eqref{quantum_kasner_beta_+_full} and \eqref{quantum_kasner_beta_-_full}. Then, regarding the transitions between the different regimes,
they coincide with the predictions of the quantum Kasner map [Eq. \eqref{trans_quantum}]. It is interesting to note that, since the first bounce occurs in sector $S_1$, the momentum $p_-$ is a constant of motion, and thus it is not modified
during this first transition. Furthermore, for the shown data, all the even bounces
take place in this sector, which is why $p_-$ only changes every two bounces.

\begin{figure}[t]
	\centering
	\begin{subfigure}[h]{0.46\textwidth}
		\centering
		\includegraphics[width=\textwidth]{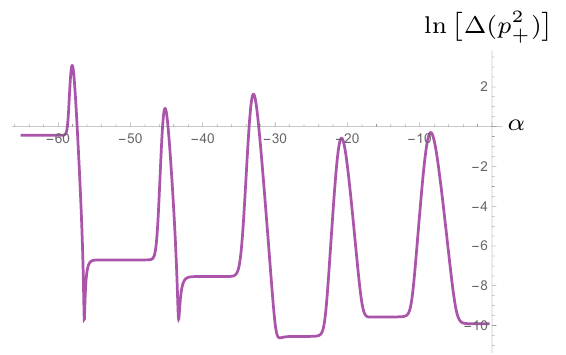}
	\end{subfigure}
	\hfill
	\begin{subfigure}[h]{0.46\textwidth}
		\centering
		\includegraphics[width=\textwidth]{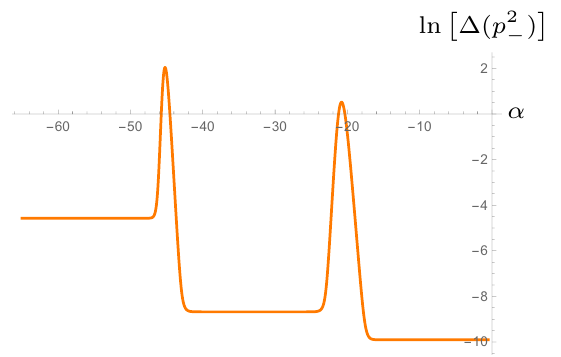}
	\end{subfigure}
	\caption{Dynamical evolution of the fluctuations $\Delta(p_+^2)$ and $\Delta(p_-^2)$ for an initial Gaussian state.}
\label{fig:moments_p_kasner}
\hfill
\begin{subfigure}[h]{\textwidth}
	\centering
	\includegraphics[width=0.46\linewidth]{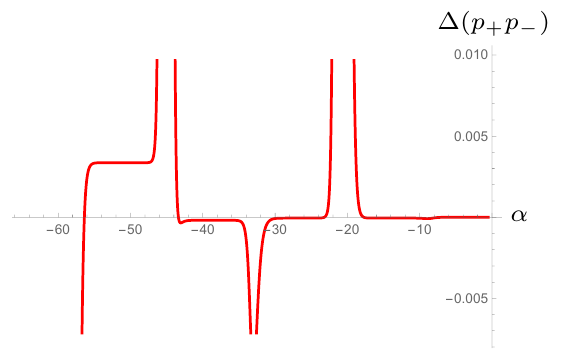}
\end{subfigure}
\caption{Dynamical evolution of the correlation $\Delta(p_+p_-)$ for an initial Gaussian state.
	The peaks correspond to the bounces against the potential walls and even if they are seemingly divergent, they are bounded.}
	\label{fig:moments_correlation_p_kasner}
\end{figure}

The numerical evolution of the quantum moments also shows the expected behavior: at each Kasner
regime, each moment $\Delta(\beta_+^ip_+^k\beta_-^kp_-^l)$ evolves as a polynomial of order $i+k$ in $\alpha$. In addition, we have checked that their transition from one Kasner epoch to the next one is in
agreement with the quantum Kasner map [Eq. \eqref{trans_quantum}]. For the purpose of illustration,
as a particular example, the evolution of the pure fluctuations $\Delta(p_+^2)$
and $\Delta(p_-^2)$ is plotted in Fig. \ref{fig:moments_p_kasner}, while the
evolution of the correlation $\Delta(p_+p_-)$ is shown in Fig. \ref{fig:moments_correlation_p_kasner}. There, one can observe
that all these variables remain constant during the different Kasner epochs, whereas their evolution shows strong
peaks at each bounce. 

\section{Conclusions}
\label{sec:conclusions}

In this paper, we have obtained the quantum Kasner
map for the Mixmaster (vacuum Bianchi IX) model within a semiclassical approximation.
For such a purpose, we have first presented the analysis of the
classical model in detail. In this context,
the Hamiltonian constraint has been deparametrized by using the (logarithm of the)
spatial volume as the time variable, and
the system has been described in terms of four variables: the shape parameters $(\beta_+,\beta_-)$, which provide
a measure of the anisotropicity of the metric, and their conjugate momenta $(p_+,p_-)$.
As is well known, due to the properties of the Bianchi IX potential,
asymptotically the dynamics of this system can be understood as
a sequence of Kasner epochs connected by quick transitions. During the
Kasner epochs, the kinetic term of the Hamiltonian dominates, and
the system follows the dynamics of a free particle. As we have argued,
the duration of each of these epochs increases as one approaches the singularity.
Furthermore,
the transitions, which happen when the potential term is not negligible,
can be properly described by the composition of the Bianchi II potential
and appropriately chosen rotations in the plane of the shape parameters.

The dynamics during the Kasner epochs can easily be solved, and it is characterized
by a set of four constant parameters. Making 
use of the constants of motion and the rotational symmetry of the system,
we have then derived the transition laws [Eq. \eqref{trans_IX}]---also known as the Kasner map---which 
relate the characteristic parameters of consecutive Kasner epochs.
In particular, the transition laws associated with the momenta $(p_+,p_-)$ can be written
in a very compact way in terms of a unique parameter $u$, as shown in Eq. \eqref{trans_u_general}.
Although this Kasner map can be iteratively applied to study $n$ consecutive transitions,
we have also provided
a closed form for the function that relates the Kasner parameters after $n$ transitions with the initial ones.

Concerning the quantum analysis of the system, we have performed a decomposition of the wave
function into its moments, in such a way that the whole physical information of the quantum state
is encoded in this infinite set of variables. We have argued that, at a semiclassical level,
it is reasonable to assume that in general the asymptotic dynamics of the system can be accurately
described by a sequence of Kasner epochs connected by quick transitions, as in the classical case.
Furthermore, we have shown that this is indeed the case in several examples where we have numerically
solved the full equations of motion.

We have then proceeded to characterize the quantum Kasner map by considering a second-order
truncation in the moments, which corresponds to neglecting terms of order $\hbar^{3/2}$ and higher,
in accordance with a semiclassical regime. We have explicitly solved the quantum dynamics in the Kasner epochs,
and have completely characterized the Kasner quantum regimes by 14 constant parameters.
Then, the constants of motion that describe the dynamics through the transitions have been obtained,
which has allowed us to construct the transition rules that provide the parameters of the postbounce
Kasner epoch in terms of those of the prebounce regime. This set of rules
defines the quantum Kasner map [Eq. \eqref{trans_quantum}] and can be considered as the most relevant result of the present paper. 

Although these rules are very long and complicated, we have been able to provide certain general features. 
For instance, for a bounce that takes place in sector $S_1$, the momentum $p_-$ is a conserved quantity of the quantum system, and thus
its transition law [Eq. \eqref{trans_general_p_-}] has the same form as its classical counterpart [Eq. \eqref{transition_p_-}].
However, the rest of the transition laws for the classical parameters [Eqs. \eqref{trans_general_p_+}--\eqref{trans_general_c_-}] 
are indeed slightly modified
due to quantum effects. Furthermore, one can notice that a postbounce constant $\widetilde{K}_{ijkl}$,
related to the moment $\Delta(\beta_+^ip_+^j\beta_-^kp_-^l)$, only depends on the prebounce
constants related to the moments of the form $\overline{\Delta(\beta_+^mp_+^n\beta_-^rp_-^s)}$,
with $m\leq i+k$, $n\leq i+j+k$, $r\leq k$, and $s\geq l$. In fact,
the fluctuation of the momentum $p_-$ is also a conserved
quantity [Eq. \eqref{trans_general_delta_p_-_2}].

An important piece of information of the full quantum Kasner map is included in the quantum
$u$-map [Eqs. \eqref{transition_u_quantum_S1}--\eqref{transition_u_quantum_S3}].
This map
takes the same form as its classical counterpart [Eq. \eqref{trans_u_general}], plus certain
correction terms linear in the variables $(v_+,v_-,r)$, which parametrize
the relative fluctuations and correlation of the momenta $(p_+,p_-)$ and thus
encode the quantum-gravity effects. Even if in this paper we have not developed
an explicit iterative analysis of this map, we have been able to point out
some features that introduce important qualitative differences with the classical model.
In particular, as explained in detail in Sec. \ref{subsec:quantum_sectors},
in the quantum model there are some exit channels (finite ranges of the parameter
$u$ or, equivalently, of the ratio $p_-/p_+$, for which the system will not have
any transition, and hence it will follow a Kasner dynamics until the singularity).
Therefore, given a set of nearby points on the phase space with their initial data in
one of these channels, all these points will follow the same qualitative evolution
and their trajectories will not diverge exponentially. In this sense, the chaotic
behavior of the classical Mixmaster model will cease. However, an open question is whether any generic
point will end up in one of these exit channels, which would imply the complete disappearance
of the chaotic repeller of the classical model.
In this respect, it is worth mentioning that in Ref. \cite{Bergeron:2017gte} the 
		instantaneous eigenvalues of the Hamiltonian have been found to be purely discrete. Even if this result might point toward the nonexistence of asymptotically free states, it is certainly not a proof, since the dynamical evolution of those eigenstates is far from trivial under a	time-dependent Hamiltonian. Therefore, this result is not in contradiction with
		our findings. Nevertheless, the existence of these states in the deep quantum regime is not clear.

Finally, in the last section, we have considered a specific quantum state that is initially
uncorrelated and obtained some features of the postbounce state by applying the
quantum Kasner map. In addition, we have numerically solved the full equations of motion
for an initial Gaussian state. This numerical analysis has served two purposes. On the one hand,
as already commented above,
we have shown that asymptotically the system indeed follows the same qualitative pattern
(a sequence of Kasner epochs connected by quick transitions) as the classical model.
On the other hand, we have tested and verified that the analytical results obtained above---in particular,
the quantum Kasner map---are accurately obeyed.

\section*{Acknowledgments}
S. F. U. is funded by an FPU fellowship of the Spanish Ministry of Universities.
We acknowledge financial support from Basque Government Grant \mbox{No.~IT956-16}
and from Grant No. FIS2017-85076-P, funded by
MCIN/AEI/10.13039/501100011033 and by “ERDF A way of making Europe.”

\appendix

\section{Relation with other parametrizations}\label{app:u_parameter}

\begin{figure}[h]
	\centering
	\includegraphics[width=0.499\linewidth]{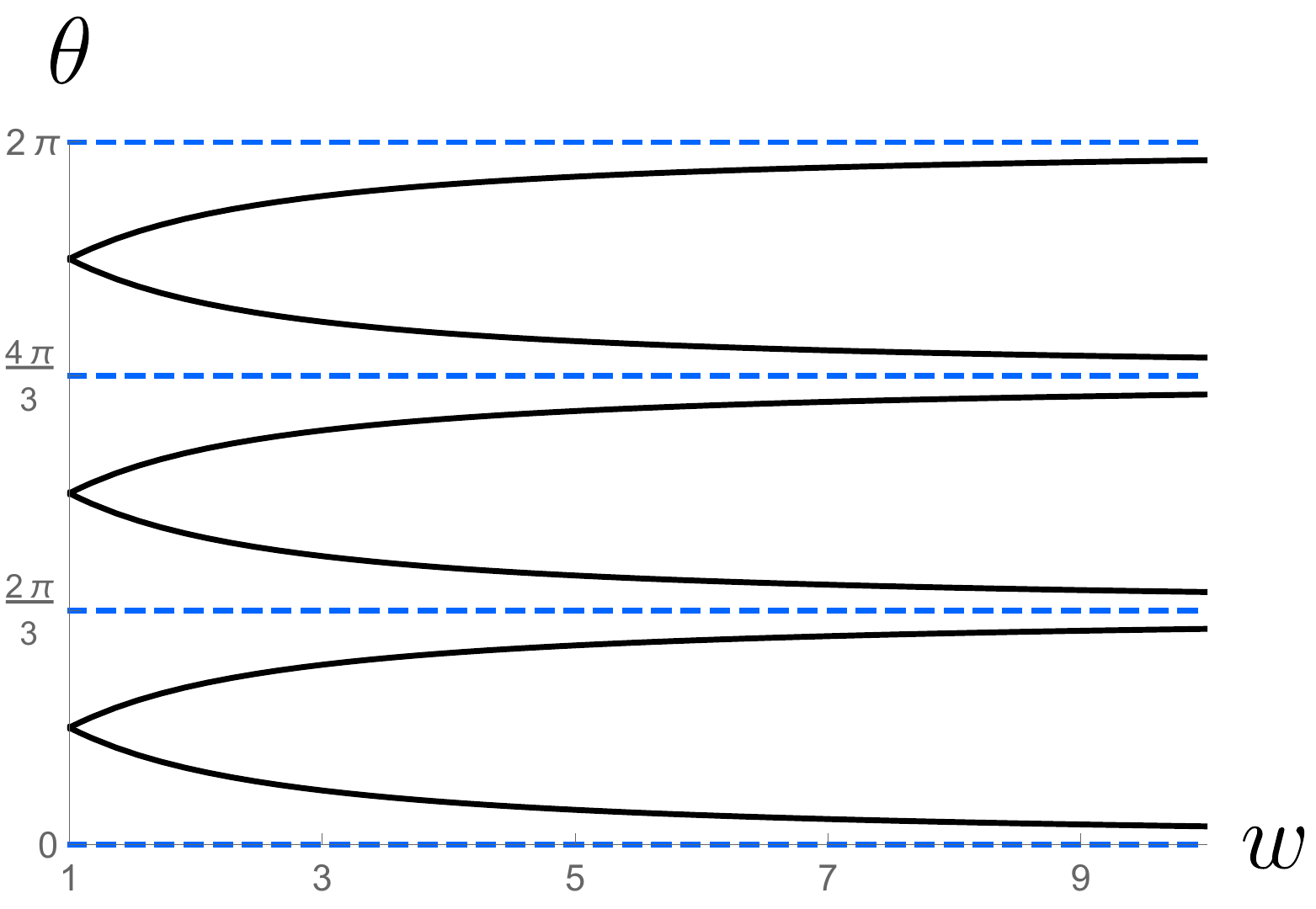}
	\caption{The black continuous curve shows the relation between the angle $\theta$ and the parameter $w$. The dashed horizontal lines are the boundaries between different the sectors $S_1$, $S_2$, and $S_3$.}
	\label{fig:theta_w}
\end{figure}

In this appendix, we provide the relation between the parametrization considered
in this paper for the angle $\theta$ [Eq. \eqref{parametrization_u}] in terms of the
parameter $u$ defined in the whole real line and another one, which we will name $w$,
that has been
widely used in the literature and is limited to $w\in[1,+\infty)$.
More precisely, the parametrization of the angle $\theta$ in terms of this parameter
$w$ is given as follows:
\begin{align}
\label{parametrization_u_literature}
\begin{split}
	\cos\left(\theta-\frac{2m(\theta)\pi}{3}\right)&=-\frac{1}{2}\left(
	\frac{1+4w+w^2}{1+w+w^2}
	\right),
	\qquad
	\left|\sin\left(\theta-\frac{2m(\theta)\pi}{3}\right)\right|=-\frac{\sqrt{3}}{2}\left(
	\frac{1-w^2}{1+w+w^2}
	\right),
\end{split}
\end{align}
where the function $m(\theta):=\left\lfloor3\theta/2\pi-1\right\rfloor$
has already been defined in the text and takes the three discrete values
$m(\theta)=0$ for $\theta\in [2\pi/3,4\pi/3)$, 
$m(\theta)=1$ for $\theta\in[4\pi/3,2\pi)$,
and $m(\theta)=-1$ for $\theta\in [0,2\pi/3)$.
This function makes the argument of the trigonometric functions lie in the interval $[2\pi/3,4\pi/3]$. Due to this fact, and the presence
of an absolute value on the sine function, the $w$ parameter is bounded to the domain $w\geq 1$.
As can be clearly seen in Fig. \ref{fig:theta_w},
since the function $\theta(w)$ from $w\in[1,\infty)$ to $\theta\in[0, 2\pi)$ 
is a surjective (1 to 6) map (and thus nonbijective), contrary to the parameter $u$ we have considered,
this parameter $w$ does not completely determine the angle $\theta$.

Combining both definitions \eqref{parametrization_u} and \eqref{parametrization_u_literature}, the relation between the two parameters can be easily derived:
\begin{align}\label{relation_w}
	w(u)=\begin{cases}
	u,&u\geq 1,
	\\
	\frac{1}{u},&u\in[0,1),
	\\
    -\frac{u+1}{u},&u\in\left[-\frac{1}{2},0\right),
	\\
	-\frac{u}{u+1},&u\in\left[-1,-\frac{1}{2}\right),
	\\
	-\frac{1}{u+1},&u\in[-2,-1),
	\\
	-u-1,&u< -2.
	\end{cases}
\end{align}
Moreover, in order to obtain the transition law for $w$, one just needs to perform the transformation $u=u(w)$ in the transition law for $u$ [Eq. \eqref{trans_u_general}].
In this way, one recovers well-known transition law,
\begin{align}\label{trans_v}
	\widetilde{w}=\begin{cases}
	\overline{w}-1,&\overline{w}\geq 2,
	\\
	\frac{1}{\overline{w}-1},&\overline{w}
	\in[1,2),
	\end{cases}
\end{align}
where $\overline{w}$ and $\widetilde{w}$ are the prebounce and postbounce values of the parameter $w$, respectively.

\vspace{0.3cm}

\section{Explicit quantum transition laws}\label{app:explicit}

In this appendix we provide the explicit transition laws for the different variables
of the system for a bounce in sector $S_1$ up to the second order in moments.
As in the main text, the tilde refers to postbounce quantities, whereas the overline
stands for prebounce objects, and in particular
$\overline{P}=(\overline p_+^2+\overline p_-^2)^{1/2}$.
\begin{align}
\label{quantum_transition_p_-}
\widetilde{p}_-&=\overline{p}_-,
\\
\label{quantum_transition_p_+}
\widetilde{p}_+&=\frac{1}{3}\left(4\overline P-5\overline p_+\right)
+\frac{2}{3\overline P^3}
\left(
\overline p_+^2 \overline{\Delta(p_-^2)}-2p_-p_+\overline{\Delta(p_+p_-)}
+p_-^2\overline{\Delta(p_+^2)}
\right),
\\
\widetilde{c}_+&=-\frac{3\overline P}{5\overline P-4\overline p_+} \overline{c}_+-\frac{2\overline P-\overline p_+}{2\left(5\overline P-4\overline p_+\right)}\ln\left[\frac{2}{3}(2 \overline p_+-\overline P)^2\right]
+\frac{12\overline p_+\overline p_-}{\overline P(5\overline P-4\overline p_+)^2}
\overline{k}_{\scriptsize_{+-}}
-\frac{12\overline p_-^2}{\overline P(5\overline P-4\overline p_+)^2}\overline{k}_{\scriptsize_{++}}
\nonumber
\\
&+\frac{1}{2\overline P^3(5\overline P-4\overline p_+)^2}
\Bigg\{
\frac{2\overline P-\overline p_+ 
}{(2 \overline p_+-\overline P)^2}
\left[21\overline p_-^4-\overline p_+^3(40\overline P-41\overline p_-)
-2\overline p_-^2\overline p_+(17\overline P-27\overline p_+)
\right] 
\nonumber
\\
&
-\frac{3\overline p_-^2\left[8\overline p_-^2+3\overline p_+(4\overline p_+-5\overline P)\right]}{2(5\overline P-4\overline p_+)}\left(8 \overline c_++\ln
\left[\frac{2}{3} (2 \overline p_+-\overline P)^2\right]\right)
\Bigg\}\overline{\Delta(p_+^2)}
\nonumber
\\
&+\frac{\overline p_-}{\overline P^3(5\overline P-4\overline p_+)^2}
\Bigg\{
\frac{3\left[2\overline p_+^2(4\overline p_+-5\overline P)+\overline p_-^2(4\overline p_++5\overline P)
	\right]}{2(5\overline P-4\overline p_+)}
\left(8 \overline c_++\ln
\left[\frac{2}{3} (2 \overline p_+-\overline P)^2\right]\right)
\nonumber
\\
&
+\frac{
1
}{(2 \overline p_+-\overline P)^2}
	\left[\overline p_+^3(68\overline P-67\overline p_+)
+2\overline p_-^2(23\overline p_+\overline P-41\overline p_+^2)-23\overline p_-^4\right]
\Bigg\}\overline{\Delta(p_+p_-)}
\nonumber
\\
&+\frac{1}{2\overline P^3(5\overline P-4\overline p_+)^2}
\Bigg[
\frac{1
}{(2 \overline p_+-\overline P)^2}
\left[\overline p_-^4(10\overline P-7\overline p_+)+\overline p_+^4(41\overline p_+-40\overline P)+2\overline p_-^2(13\overline p_+^3-4\overline P\overline p_+^2)\right]
\nonumber
\\
&
-\frac{3\overline p_+(10\overline p_-^2\overline P-5\overline p_+^2\overline P+4\overline p_+^3)}{2(5\overline P-4\overline p_+)}
\left(8 \overline c_++\ln
\left[\frac{2}{3} (2 \overline p_+-\overline P)^2\right]\right)
\Bigg]\overline{\Delta(p_-^2)},
\label{quantum_transition_c_+}
\\
\widetilde{c}_-&=c_-+\frac{4\overline p_-}{5\overline P-4\overline p_+} \overline{c}_++\frac{\overline p_-}{2\left(5\overline P-4\overline p_+\right)}\ln\left[\frac{2}{3}(2 \overline p_+-\overline P)^2\right]
+\frac{4\overline p_- (4\overline P-5\overline p_+)}{\overline P(5\overline P-4\overline p_+)^2}\overline{k}_{\scriptsize_{++}}-
\frac{4\overline p_+(4\overline P-5\overline p_+)}{\overline P(5\overline P-4\overline p_+)^2}
\overline{k}_{\scriptsize_{+-}}
\nonumber
\\
&
+\frac{\overline p_-}{2\overline P^3(5\overline P-4\overline p_+)^2}
\Bigg\{\frac{5\overline p_+\left[8\overline p_-^2-3\overline p_+(5\overline P-4\overline p_+)\right]}{2(5\overline P-4\overline p_+)}
\left(8 \overline c_++\ln
\left[\frac{2}{3} (2 \overline p_+-\overline P)^2\right]\right)
\nonumber
\\
&
-
\frac{1   
}{(2 \overline p_+-\overline P)^2}
\left[5\overline p_-^4+2\overline p_-^2\overline p_+(2\overline P-13\overline p_+)
+3\overline p_+^3(14\overline P-13\overline p_+
) \right]
\Bigg\}\overline{\Delta(p_-^2)}
\nonumber
\\
&+\frac{1}{\overline P^3(5\overline P-4\overline p_+)^2}
\Bigg\{
\frac{34\overline p_+\overline p_-^2\overline P-20\overline p_-^4
	-\overline p_+^3(41\overline P-40\overline p_+)}{2(5\overline P-4\overline p_+)}\left(8 \overline c_++\ln
\left[\frac{2}{3} (2 \overline p_+-\overline P)^2\right]\right)
\nonumber
\\
&
+\frac{2\overline P-\overline p_+
}{(2 \overline p_+-\overline P)^2}
\left[7\overline p_-^4+\overline p_+^3(14\overline P-13\overline p_+)
-2\overline p_+\overline p_-^2(4\overline P-\overline p_+)\right]
\Bigg\}\overline{\Delta(p_+p_-)}
\nonumber
\\
&+\frac{\overline p_-}{2\overline P^3(5\overline P-4\overline p_+)^2}
\Bigg\{
\frac{\left[\overline p_-^2(7\overline P-60\overline p_+)
	+2\overline p_+^2(41\overline P-40\overline p_+)
	\right]}{2(5\overline P-4\overline p_+)}
\left(8 \overline c_++\ln
\left[\frac{2}{3} (2 \overline p_+-\overline P)^2\right]\right)
\nonumber
\\
&
+\frac{1
}{(2 \overline p_+-\overline P)^2}
\left[
3\overline p_+^3(40\overline P-41\overline p_+)
+2\overline p_+\overline p_-^2(41\overline P-73\overline p_+)-31\overline p_-^4
\right]
\Bigg\}\overline{\Delta(p_+^2)},
\label{quantum_transition_c_-}
\\
\widetilde{\Delta(p_-^2)}&=\overline{\Delta(p_-^2)},
\label{quantum_transition_delta_p_-_2}
\\
\widetilde{\Delta(p_+^2)}&=\frac{1}{9\overline P^2}
\left[16\overline p_-^2
\overline{\Delta(p_-^2)}-8\overline p_-(5\overline P-4\overline p_+)
\overline{\Delta(p_+p_-)}+(5\overline P-4\overline p_+)^2
\overline{\Delta(p_+^2)}
\right],
\label{quantum_transition_delta_p_+_2}
\\
\widetilde{\Delta(p_+p_-)}&=\frac{1}{3\overline P}
\left[4\overline p_-
\overline{\Delta(p_-^2)}-(5\overline P-4\overline p_+)
\overline{\Delta(p_+p_-)}
\right],
\label{quantum_transition_delta_p_+_p_-}
\end{align}
\begin{align}
\widetilde{k}_{\scriptsize_{+-}}&=-\frac{3\overline{P}}{5\overline P-4\overline p_+}\overline{k}_{\scriptsize_{+-}}
+\frac{\overline p_-}{\overline P(5\overline P-4\overline p_+)}\left[
\frac{2\overline P-\overline p_+}{2\overline p_+-\overline P}
+\frac{3\overline p_+}{2(5\overline P-4\overline p_+)}
\left(8 \overline c_++\ln
\left[\frac{2}{3} (2 \overline p_+-\overline P)^2\right]\right)
\right]\overline{\Delta(p_-^2)}
\nonumber
\\
&
-\frac{1}{\overline P(5\overline P-4\overline p_+)}\left[
\frac{(2\overline P-\overline p_+)^2}{2\overline p_+-\overline P}
+\frac{3\overline p_-^2}{2(5\overline P-4\overline p_+)}
\left(8 \overline c_++\ln
\left[\frac{2}{3} (2 \overline p_+-\overline P)^2\right]\right)
\right]\overline{\Delta(p_+p_-)},
\label{quantum_transition_delta_beta_+_p_-}
\\
\widetilde{k}_{\scriptsize_{--}}&=\overline{k}_{\scriptsize_{--}}
+\frac{4\overline p_-}{5\overline P-4\overline p_+}
\overline{k}_{\scriptsize_{+-}}
-\frac{1}{\overline P(5\overline P-4\overline p_+)}\left[
\frac{\overline p_-^2}{2\overline p_+-\overline P}
+\frac{\overline p_+(4\overline P-5\overline p_+)}{2(5\overline P-4\overline p_+)}
\left(8 \overline c_++\ln
\left[\frac{2}{3} (2 \overline p_+-\overline P)^2\right]\right)
\right]\overline{\Delta(p_-^2)}
\nonumber
\\
&
+\frac{\overline p_-}{\overline P(5\overline P-4\overline p_+)}\left[
\frac{2\overline P-\overline p_+}{2\overline p_+-\overline P}
+\frac{(4\overline P-5\overline p_+)}{2(5\overline P-4\overline p_+)}
\left(8 \overline c_++\ln
\left[\frac{2}{3} (2 \overline p_+-\overline P)^2\right]\right)
\right]\overline{\Delta(p_+p_-)}
,
\label{quantum_transition_delta_beta_-_p_-}
\\
\widetilde{k}_{\scriptsize_{++}}&=\overline{k}_{\scriptsize_{++}}
-\frac{4\overline p_-}{5\overline P-4\overline p_+}
\overline{k}_{\scriptsize_{+-}}
+\frac{1}{\overline P^2}\left[
\frac{(2\overline P-\overline p_+)^2}{3(2\overline p_+-\overline P)}
+\frac{\overline p_-^2}{2(5\overline P-4\overline p_+)}
\left(8 \overline c_++\ln
\left[\frac{2}{3} (2 \overline p_+-\overline P)^2\right]\right)
\right]\overline{\Delta(p_+^2)}
\nonumber
\\
&
-\frac{\overline{\Delta(p_+p_-)}\overline p_-}{\overline P^2(5\overline P-4\overline p_+)}\left[
\frac{(13\overline P-8\overline p_+)(2\overline P-\overline p_+)}{3(2\overline p_+-\overline P)}
+\frac{(4\overline p_-^2+5\overline p_+ \overline P -4\overline p_+^2)}{2(5\overline P-4\overline p_+)}
\left(8 \overline c_++\ln
\left[\frac{2}{3} (2 \overline p_+-\overline P)^2\right]\right)
\right]
\nonumber
\\
&
+\frac{2\overline p_-^2}{\overline P^2(5\overline P-4\overline p_+)}\left[
\frac{2(2\overline P-\overline p_+)}{3(2\overline p_+-\overline P)}
+\frac{\overline p_+}{5\overline P-4\overline p_+}
\left(8 \overline c_++\ln
\left[\frac{2}{3} (2 \overline p_+-\overline P)^2\right]\right)
\right]\overline{\Delta(p_-^2)},
\label{quantum_transition_delta_beta_+_p_+}
\\
\widetilde{k}_{\scriptsize_{-+}}&=
\frac{4\overline p_-}{3\overline P}
\overline{k}_{\scriptsize_{--}}
-\frac{{5\overline P-4\overline p_+}}{3\overline P}
\overline{k}_{\scriptsize_{-+}}
+	\frac{16\overline p_-^2}{3\overline P(5\overline P-4\overline p_+)}
\overline{k}_{\scriptsize_{+-}}
-\frac{4\overline p_-}{3\overline P}
\overline{k}_{\scriptsize_{++}}
\nonumber
\\
&
+\frac{\overline{\Delta(p_+p_-)}}{\overline 3P^2(5\overline P-4\overline p_+)}\left[
\frac{(13\overline P-8\overline p_+)\overline p_-^2}{2\overline p_+-\overline P}
+\frac{(16\overline P^3-57 \overline P \overline p_+^2+40\overline p_+^3)}{2(5\overline P-4\overline p_+)}
\left(8 \overline c_++\ln
\left[\frac{2}{3} (2 \overline p_+-\overline P)^2\right]\right)
\right]
\nonumber
\\
&
-\frac{2\overline p_-}{3\overline P^2(5\overline P-4\overline p_+)}\left[
\frac{2\overline p_-^2}{2\overline p_+-\overline P}
+\frac{\overline p_+(4\overline P-5\overline p_+)}{5\overline P-4\overline p_+}
\left(8 \overline c_++\ln
\left[\frac{2}{3} (2 \overline p_+-\overline P)^2\right]\right)
\right]\overline{\Delta(p_-^2)}
\nonumber
\\
&
-\frac{\overline p_-}{3\overline P^2}\left[
\frac{2\overline P-\overline p_+}{2\overline p_+-\overline P}
+\frac{(4\overline P-5\overline p_+)}{2(5\overline P-4\overline p_+)}
\left(8 \overline c_++\ln
\left[\frac{2}{3} (2 \overline p_+-\overline P)^2\right]\right)
\right]\overline{\Delta(p_+^2)},
\label{quantum_transition_delta_beta_-_p_+}
\\
\widetilde{c}_{\scriptsize_{++}}&=\frac{9\overline P^2}{(5 \overline P-4 \overline p_+)^2}\overline{c}_{\scriptsize_{++}}
+\frac{1}{(5 \overline P-4 \overline p_+)^2} 
\left[\frac{9 
	\overline p_-^2}{5 \overline P-4 \overline p_+}+\frac{6 (2 \overline P-\overline p_+)^2}{2 \overline p_+-\overline P}
\left(8\overline c_++\ln \left[\frac{2}{3} (2 \overline p_+-\overline P)^2\right]\right)
\right]\overline{k}_{\scriptsize_{++}}
\nonumber
\\
\nonumber
&
-\frac{\overline p_- }{(5 \overline P-4 \overline p_+)^2} \left[\frac{6(2 \overline P-\overline p_+)}{2 \overline p_+-\overline P}+\frac{9
	\overline p_+}{(5 \overline P-4
	\overline p_+)^3}
\left(8\overline c_++\ln \left[\frac{2}{3} (2 \overline p_+-\overline P)^2\right]\right)
\right]\overline{k}_{\scriptsize_{+-}}
\\
\nonumber
&+\frac{1}{\overline P^2(5 \overline P-4 \overline p_+)^2}
\left[
\frac{(2\overline P-\overline p_+)^2}{2\overline p_+-\overline P}
+\frac{3\overline p_-^2}{2(5 \overline P-4 \overline p_+) }\left(8\overline c_++\ln \left[\frac{2}{3} (2 \overline p_+-\overline P)^2\right]\right)	
\right]^2\overline{\Delta(p_+^2)} 
\\
\nonumber
&-\frac{\overline p_- }{ \overline P^2(5 \overline P-4 \overline p_+)^2} \Bigg\{ \frac{9 \overline p_+ \overline p_-^2
	}{2(5 \overline P-4 \overline p_+)^2}
\left(8\overline c_++\ln \left[\frac{2}{3} (2 \overline p_+-\overline P)^2\right]\right)^2
+\frac{2(2
	\overline P-\overline p_+)^3}{(2 \overline p_+-\overline P)^2}
\\\nonumber
&
+\frac{3 (2 \overline P-\overline p_+)
	\left[\overline p_-^2+(2 \overline P-\overline p_+) \overline p_+\right]}{(5 \overline P-4 \overline p_+) (2 \overline p_+-\overline P)}
\left(8\overline c_++\ln \left[\frac{2}{3} (2 \overline p_+-\overline P)^2\right]\right)
\Bigg\}\overline{\Delta(p_+p_-)}
\\
&	+\frac{\overline p_-^2}{\overline P^2(5 \overline P-4 \overline p_+)^2}
\left[
\frac{2\overline P-\overline p_+}{2\overline p_+-\overline P}
+\frac{3\overline p_+}{2(5 \overline P-4 \overline p_+) }
\left(8\overline c_++\ln \left[\frac{2}{3} (2 \overline p_+-\overline P)^2\right]\right)	
\right]^2\overline{\Delta(p_-^2)} ,
\label{quantum_transition_delta_beta_+_2}
\\
\widetilde{c}_{\scriptsize_{--}}&=\overline{c}_{\scriptsize_{--}}+
\frac{16\overline{p}_-^2}{(5 \overline P-4 \overline p_+)^2}
\overline{c}_{\scriptsize_{++}}
+
\frac{8\overline{p}_-}{5 \overline P-4 \overline p_+}
\overline{c}_{\scriptsize_{+-}}
\nonumber
\\
&
+\frac{\overline p_-^2}{\overline P^2(5 \overline P-4 \overline p_+)^2}
\left[
\frac{2\overline P-\overline p_+}{2\overline p_+-\overline P}
+\frac{(4\overline P-5\overline p_+)}{2(5 \overline P-4 \overline p_+) }	
\left(8\overline c_++\ln \left[\frac{2}{3} (2 \overline p_+-\overline P)^2\right]\right)
\right]^2\overline{\Delta(p_+^2)}
\nonumber
\end{align}
\begin{align}
&
+\frac{1}{\overline P^2(5\overline P-4\overline p_+)^2}\left[
\frac{\overline p_-^2}{2\overline p_+-\overline P}
+\frac{\overline p_+(4\overline P-5\overline p_+)}{2(5\overline P-4\overline p_+)}
\left(8\overline c_++\ln \left[\frac{2}{3} (2 \overline p_+-\overline P)^2\right]\right)
\right]^2\overline{\Delta(p_-^2)}
\nonumber
\\
\nonumber
&+\frac{4\overline{p}_-^2}{\overline P(5 \overline P-4 \overline p_+)^2}\left[\frac{2
	(2 \overline P-\overline p_+)}{2 \overline p_+-\overline P}+\frac{
	(4 \overline P-5 \overline p_+)}{ 5 \overline P-4 \overline p_+}
 \left(8 c_++\ln \left[\frac{2}{3} (2
\overline p_+-\overline P)^2\right]\right)
\right]
\overline{k}_{\scriptsize_{++}}
\\
\nonumber
&+\frac{\overline{p}_-}{\overline P(5 \overline P-4 \overline p_+)}\left[\frac{2
	(2 \overline P-\overline p_+)}{2 \overline p_+-\overline P}+\frac{
	(4 \overline P-5 \overline p_+)}{ 5 \overline P-4 \overline p_+}
\left(8 c_++\ln \left[\frac{2}{3} (2
\overline p_+-\overline P)^2\right]\right)
\right]
\overline{k}_{\scriptsize_{-+}}
\\
\nonumber
&-\frac{4\overline{p}_-}{\overline P(5 \overline P-4 \overline p_+)^2}\left[\frac{2
	\overline p_-^2}{2 \overline p_+-\overline P}+\frac{
	\overline p_+(4 \overline P-5 \overline p_+)}{ 5 \overline P-4 \overline p_+}
\left(8 c_++\ln \left[\frac{2}{3} (2
\overline p_+-\overline P)^2\right]\right)
\right]
\overline{k}_{\scriptsize_{+-}}
\\
\nonumber
&-\frac{1}{\overline P(5 \overline P-4 \overline p_+)}\left[\frac{2
	\overline p_-^2}{2 \overline p_+-\overline P}+\frac{
	\overline p_+(4 \overline P-5 \overline p_+) }{ 5 \overline P-4 \overline p_+}
\left(8 c_++\ln \left[\frac{2}{3} (2
\overline p_+-\overline P)^2\right]\right)
\right]
\overline{k}_{\scriptsize_{--}}
\\
&-\frac{\overline{p}_-}{\overline P^2(5 \overline P-4 \overline p_+)^2}
\Bigg\{\frac{\left[\overline p_-^2 (4 \overline P+3 \overline p_+)+\overline p_+^2 (13 \overline p_+-14
	\overline P)\right]}{ (5 \overline P-4 \overline p_+) (2
	\overline p_+-\overline P)} \left(8 \overline c_++\ln \left[\frac{2}{3} (2\overline p_+-\overline P)^2\right]\right)+\frac{2\overline p_-^2 (2 \overline P-\overline p_+)}{(2 \overline p_+-\overline P)^2}
\nonumber
\\
&
+\frac{\overline p_+ (4 \overline P-5 \overline p_+)^2 }{2 (5 \overline P-4
	\overline p_+)^2}\left(8 \overline c_++\ln
\left[\frac{2}{3} (2 \overline p_+-\overline P)^2\right]\right)^2
\Bigg\}\overline{\Delta(p_+p_-)},
\label{quantum_transition_delta_beta_-_2}
\\
\nonumber
\widetilde{c}_{\scriptsize_{+-}}&=-
\frac{12\overline p_-\overline{P}}{(5 \overline P-4 \overline p_+)^2}\overline{c}_{\scriptsize_{++}}-\frac{3\overline{P}}{5 \overline P-4 \overline p_+}
\overline{c}_{\scriptsize_{+-}}
\\
\nonumber
&
-\frac{\overline p_-}{\overline P(5 \overline P-4 \overline p_+)^2}
\Bigg\{
\frac{3\left[8\overline p_-^2-\overline p_+(5\overline P-4\overline p_+)\right] }{2 (5 \overline P-4 \overline p_+)}\left(8 \overline c_++\ln \left[\frac{2}{3} (2\overline p_+-\overline P)^2\right]\right)
\\
\nonumber
&
+\frac{22\overline p_-^2+\overline p_+(26\overline p_+-19\overline P)}{2 \overline p_+-\overline P}\Bigg\}
\overline{k}_{\scriptsize_{++}}
\\
&-\frac{1}{\overline P(5 \overline P-4 \overline p_+)}
\left[
\frac{3 \overline p_-^2 }{2 (5 \overline P-4 \overline p_+)}
\left(8 \overline c_++\ln \left[\frac{2}{3} (\overline P-2
\overline p_+)^2\right]\right)
+\frac{(2
	\overline P-\overline p_+)^2}{2 \overline p_+-\overline P}\right]\overline{k}_{\scriptsize_{-+}}
\nonumber
\\
\nonumber
&
+\frac{\overline{k}_{\scriptsize_{+-}}}{\overline P(5 \overline P-4 \overline p_+)^2}
\Bigg\{
\frac{3\overline p_+\left[8\overline p_-^2-\overline p_+(5\overline P-4\overline p_+)\right] }{2 (5 \overline P-4 \overline p_+)}\left(8 \overline c_++\ln \left[\frac{2}{3} (2\overline p_+-\overline P)^2\right]\right)
+\frac{\overline p_-^2(11\overline P-4\overline p_+)}{2 \overline p_+-\overline P}\Bigg\}
\\
\nonumber
&	+\frac{\overline p_-}{\overline P(5 \overline P-4 \overline p_+)}
\left[
\frac{3\overline p_+}{2 (5 \overline P-4 \overline p_+)}\left(8 \overline c_++\ln \left[\frac{2}{3} (\overline P-2
\overline p_+)^2\right]\right)
+\frac{2\overline P-\overline p_+}{2 \overline p_+-\overline P}\right]
\overline{k}_{\scriptsize_{--}}
\\\nonumber
&-\frac{\overline p_-}{\overline P^2(5 \overline P-4 \overline p_+)^2}
\Bigg\{
\frac{(2 \overline P-\overline p_+)\left[11\overline p_-^2+\overline p_+(13\overline p_+-14\overline P)\right]
}{2(5 \overline P-4 \overline p_+)(2\overline p_+-\overline P)}	\left(8 \overline c_++\ln \left[\frac{2}{3} (2\overline p_+-\overline P)^2\right]\right)
\\
\nonumber
&
+\frac{(2 \overline P-\overline p_+)^3}{(2\overline p_+-\overline P)^2}
+\frac{3\overline p_-^2(4\overline P-5\overline p_+) 
}{4 (5 \overline P-4 \overline p_+)^2}
	\left(8 \overline c_++\ln \left[\frac{2}{3} (2\overline p_+-\overline P)^2\right]\right)^2
\Bigg\}\overline{\Delta(p_+^2)}
\\
\nonumber
&-\frac{\overline p_-}{\overline P^2(5 \overline P-4 \overline p_+)^2}
\Bigg\{
\frac{\overline p_-^2(2\overline P-\overline p_+)}{(2\overline p_+-\overline P)^2}+\frac{3\overline p_+^2(4\overline P-5\overline p_+)
}{4 (5 \overline P-4 \overline p_+)^2}
 \left(8 \overline c_++\ln \left[\frac{2}{3} (2\overline p_+-\overline P)^2\right]\right)^2
\\
\nonumber
&
+\frac{\overline p_+\left[11\overline p_-^2+\overline p_+(13\overline p_+-14\overline P)\right]
}{2(5 \overline P-4 \overline p_+)(2\overline p_+-\overline P)}
	\left(8 \overline c_++\ln \left[\frac{2}{3} (2\overline p_+-\overline P)^2\right]\right)
\Bigg\}\overline{\Delta(p_-^2)}
\\\nonumber
&+\frac{1}{\overline P^2(5 \overline P-4 \overline p_+)^2}
\Bigg\{
\frac{(2 \overline P-\overline p_+)^2}{(2 \overline p_+-\overline P)^2}+
\frac{3 \overline p_-^2\overline p_+(4 \overline P-5 \overline p_+) }{2 (5 \overline P-4 \overline p_+)^2}
\left(8 \overline c_++\ln \left[\frac{2}{3} (2\overline p_+-\overline P)^2\right]\right)^2
\\
\label{quantum_transition_delta_beta_+_beta_-}
&
+\frac{\left[
	11\overline p_-^4+2\overline p_-\overline p_+
	(13\overline p_+-4\overline P)+\overline p_+^3
	(41\overline p_+-40\overline P)
	\right]}{2 (5 \overline P-4 \overline p_+)
	(2 \overline p_+-\overline P)}
\left(8 \overline c_++\ln
\left[\frac{2}{3} (2\overline p_+-\overline P)^2\right]\right)
\Bigg\}\overline{\Delta(p_+p_-)}.
\end{align}

\end{document}